\documentclass[preprint]{elsarticle}

\usepackage{color}
\usepackage{graphicx}
\usepackage{tikz,tikz-3dplot}

\usepackage{amsmath}
\usepackage{amsfonts}
\usepackage{amssymb}
\usepackage{amsthm}
\usepackage{epstopdf}
\usepackage{algorithmic}
\usepackage{setspace}
\usepackage[cm]{fullpage}
\usepackage{multirow}
\usepackage{url}

\makeatletter  
\renewcommand{\verbatim@font}{%
  \ttfamily\small\catcode`\<=\active\catcode`\>=\active%
}
\makeatother   

\newcommand\restr[2]{{
  \left.\kern-\nulldelimiterspace 
  #1 
  \vphantom{\big|} 
  \right|_{#2} 
}}

\newcommand{\comment}[1]{}

\newcommand{\vdotss}[1]{
\begin{tikzpicture}[scale=#1]
\draw [fill] (0,0) circle(0.01);
\draw [fill] (0,0.1) circle(0.01);
\draw [fill] (0,0.2) circle(0.01);
\draw [fill] (0,0.3) circle(0);
\end{tikzpicture}%
}

\theoremstyle{remark}

\newtheorem{algorithm}{Algorithm}

\newtheorem{example}{Example}
\newenvironment{acknow}[1]{\vspace*{#1mm}\noindent\bf Acknowledgements\;\rm}

\newcommand{\nedelec}{N\'ed\'elec }

\newcommand{\R}{\mathbb{R}}

\newcommand{\up}[1]{\overline{#1}}							

\def\stiff{\textrm{\textbf{K}}}
\def\mass{\textrm{\textbf{M}}}

\def\Rd{{\mathbb R}^d}
\def\Norms#1{\left\Vert#1\right\Vert}
\def\O{\Omega}
\def\CO{C_F}
\def\Ieff{I_{\textrm{eff}}}

\def\Ltwo{\mathit{L}^2}				
\def\H{\mathit{H}}					
\newcommand{\Hoz}{\mathring{\H}^1(\O)}
\newcommand{\Hoc}{\mathring{\H}_{\Gamma_D}(\curl,\O)}

\def\dvg{\textrm{div}}
\def\curl{\textrm{curl}}
\def\curlv{\underline{\textrm{curl}}}
\def\dx{{\rm dx}}
\def\xh{\hat{x}}
\def\dxh{{\rm d\xh}}
\def\sh{\hat{s}}
\def\dsh{{\rm d\sh}}
\def\etah{\hat{\eta}}
\def\rt{{\rm RT}}
\def\ned{{\rm Ned}}
\newcommand{\abs}[1]{\left|#1\right|}
\newcommand{\sign}[1]{[{\rm sign}_K^{#1}]}
\newcommand{\signs}[2]{[{\rm sign}_{#1}^{#2}]}

\def\nodes{{\mathcal N}}
\def\edges{{\mathcal E}}
\def\faces{{\mathcal F}}
\def\elems{{\mathcal T}}

\journal{Applied Mathematics and Computation}

\begin{document}

\begin{frontmatter}

\title{Fast MATLAB assembly of FEM matrices in 2D and 3D: Edge elements}

\author[jyu]{I.~Anjam\corref{cor}}
\ead{immanuel.anjam@jyu.fi}

\author[bohemia,ascr]{J.~Valdman}
\ead{jvaldman@prf.jcu.cz}

\cortext[cor]{Corresponding author at: P.O.Box 35 (Agora), FI-40014 University of Jyv\"askyl\"a, Finland}
\address[jyu]{Department of Mathematical Information Technology, University of Jyv\"askyl\"a, Finland}
\address[bohemia]{Institute of Mathematics and Biomathematics, University of South Bohemia, \v Cesk\'e Bud\v ejovice, Czech Republic}
\address[ascr]{Institute of Information Theory and Automation of the ASCR, Prague, Czech Republic}

\begin{abstract}
We propose an effective and flexible way to assemble finite element stiffness and mass matrices in MATLAB. We apply this for problems discretized by edge finite elements. Typical edge finite elements are Raviart-Thomas elements used in discretizations of $\H(\dvg)$ spaces and \nedelec elements in discretizations of $\H(\curl)$ spaces. We explain vectorization ideas and comment on a freely available MATLAB code which is fast and scalable with respect to time.
\end{abstract}


\end{frontmatter}


\section{Introduction}

Elliptic problems containing the full gradient operator $\nabla$ of scalar or vector arguments are formulated in weak forms in $\H^1$ Sobolev spaces and discretized using nodal finite element functions. Efficient MATLAB vectorization of the assembly routine of stiffness matrices for the linear nodal finite element was explained by T. Rahman and J. Valdman in  \cite{RaVa}. The focus of this paper is generalizing the ideas of \cite{RaVa} to arbitrary finite elements, including higher order elements, and vector problems operating with the divergence operator $\dvg$ and the rotation operator $\curl$. Such problems appear in electromagnetism and are also related to various mixed or dual problems in mechanics. Weak forms of these problems are defined in $\H(\dvg)$ and $\H(\curl)$ Sobolev spaces. A finite element discretization is done in terms of edge elements, typically Raviart-Thomas elements \cite{RT} for $\H(\dvg)$ problems and \nedelec elements \cite{Ned} for $\H(\curl)$ problems. Edge element basis functions are not defined on the nodes of 2D triangular or 3D tetrahedral meshes, but on edges and faces. Edge elements provide only partial continuity over element boundaries: continuity of normal vector component for $\H(\dvg)$ problems and continuity of tangential vector component for $\H(\curl)$ problems.

The method of finite elements applied to $\H(\dvg)$ and $\H(\curl)$ problems and its implementation has been well documented, see for instance \cite{So} including higher order polynomials defined through hierarchical bases. A user can find many software codes (for instance NGSOLVE \cite{Sch} or HERMES \cite{So2}) written in object oriented languages allowing for higher order  elements defined on elements with curved boundaries. These codes are very powerful, capable of high complexity computations and they provide certain flexibility via user interface. However, such codes are not so easy to understand and modify unless one is quite familiar with the code. We believe that our MATLAB code is more convenient for students and researchers who wish to become familiar with edge elements and prefer to have their own implementation. We consider the lowest order linear edge elements defined on 2D triangles and 3D tetrahedra only. However, it is straightforward to extend the code to use higher order elements, since the assembly routines remain almost the same regardless of the element order.


There are plenty of papers \cite{Chen,FuPrWi,HaJu,CuJaSc} dedicated to implementing vectorized FEM assembly routines for nodal elements in MATLAB. In \cite{HaJu} the authors also discuss the Raviart-Thomas element in 3D, but do not provide the program code. The iFEM package \cite{Chen} has efficient implementation of FEM assembly routines for various different linear and higher order elements. The paper \cite{BaCa} considers the implementation of Raviart-Thomas elements (in a non-vectorized way), providing a good inspiration for the implementation of a multigrid based solver for $\H(\dvg)$ majorant minimization \cite{Va} by the second author.

Our implementation generalizes the approach of \cite{RaVa} to work with arbitrary affine finite elements. It is based again on operations with long vectors and arrays in MATLAB and it is  reasonably scalable for large size problems. On a typical computer with a decent processor and enough system memory, the 2D/3D assemblies of FEM matrices are very fast. For example, a 2D assembly of matrices with around 10 million rows takes less than a minute. Vectorization of calculations typically requires  more system memory, but the performance degrades only when the system memory becomes full. The software described in this paper is available for download at MATLAB Central at 
\begin{center}
\url{http://www.mathworks.com/matlabcentral/fileexchange/46635} 
\end{center}
It also includes an implementation of the linear nodal finite element in 2D and 3D with the generalized approach, so an interested reader can compare the performance of the code described in \cite{RaVa}. The key idea in our generalized approach is to vectorize the integration procedure of scalar and vector valued functions on affine meshes. This allows also for fast evaluation of norms of functions.

The paper is divided as follows: In Section 2 we briefly describe the implemented linear edge elements. In Section 3 we go through the particular constructions related to the implementation of the elements and vectorization details. We also show the performance of the vectorized assembly routines with respect to time and scalability. Section 4 illustrates two applications of edge elements: a functional majorant minimization in a posteriori error analysis and solving of an electromagnetic problem.


\section{Linear edge elements}

\tdplotsetmaincoords{60}{120}
\begin{figure} [b]
\centering
\mbox{
\begin{tikzpicture}[scale=1.4]
\draw (0.6,-0.6) node [] {\footnotesize 2D Raviart-Thomas};	
\draw (0.6,-0.9) node [] {\footnotesize (edges, \#dof=3)};
\draw [->] (0,0) -- (1.3,0);						
\draw [->] (0,0) -- (0,1.3);
\draw (1,0) -- (0,1);							
\draw [fill,color=black] (0,0) circle(0.03);			
\draw [fill,color=black] (1,0) circle(0.03);
\draw [fill,color=black] (0,1) circle(0.03);
\draw [color=gray] (1.45,-0.1) node [] {$\xh_1$};		
\draw [color=gray] (-0.19,1.35) node [] {$\xh_2$};
\draw [thick,->] (0.5,-0.05) -- (0.5,-0.35);			
\draw [thick,->] (-0.05,0.5) -- (-0.35,0.5);
\draw [thick,->] (0.54,0.54) -- (0.76,0.76);
\draw [color=gray] (0.65,-0.3) node [] {\footnotesize $3$};		
\draw [color=gray] (1.15,0.55) node [] {\footnotesize $1: \hat{e}_1, \hat{n}_1$};
\draw [color=gray] (-0.3,0.7) node [] {\footnotesize $2$};
\end{tikzpicture}
}
\mbox{
\begin{tikzpicture}[scale=1.4]
\draw (0.6,-0.6) node [] {\footnotesize 2D \nedelec};	
\draw (0.6,-0.9) node [] {\footnotesize (edges, \#dof=3)};
\draw [->] (0,0) -- (1.3,0);						
\draw [->] (0,0) -- (0,1.3);
\draw (1,0) -- (0,1);							
\draw [fill,color=black] (0,0) circle(0.03);			
\draw [fill,color=black] (1,0) circle(0.03);
\draw [fill,color=black] (0,1) circle(0.03);
\draw [color=gray] (1.45,-0.1) node [] {$\xh_1$};		
\draw [color=gray] (-0.19,1.35) node [] {$\xh_2$};
\draw [thick,->] (0.26,-0.1) -- (0.69,-0.1);			
\draw [thick,->] (-0.1,0.69) -- (-0.1,0.26);
\draw [thick,->] (0.72,0.42) -- (0.42,0.72);
\draw [color=gray] (0.65,-0.3) node [] {\footnotesize $3$};		
\draw [color=gray] (0.9,0.85) node [] {\footnotesize $1: \hat{e}_1, \hat{t}_1$};
\draw [color=gray] (-0.28,0.3) node [] {\footnotesize $2$};
\end{tikzpicture}
}
\mbox{
\begin{tikzpicture}[scale=1.3,tdplot_main_coords]
\draw (0,0.3,-1.15) node [] {\footnotesize 3D Raviart-Thomas};	
\draw (0,0.3,-1.5) node [] {\footnotesize (faces, \#dof=4)};
\draw [->] (0,0,0) -- (1.5,0,0);						
\draw [->] (0,0,0) -- (0,1.5,0);
\draw [->] (0,0,0) -- (0,0,1.4);
\draw (1,0,0) -- (0,1,0);							
\draw (0,1,0) -- (0,0,1);
\draw (0,0,1) -- (1,0,0);
\draw [fill,color=black] (0,0,0) circle(0.04);				
\draw [fill,color=black] (1,0,0) circle(0.04);
\draw [fill,color=black] (0,1,0) circle(0.04);
\draw [fill,color=black] (0,0,1) circle(0.04);
\draw [color=gray] (1.6,-0.1,0) node [] {$\xh_1$};			
\draw [color=gray] (0,1.7,0) node [] {$\xh_2$};
\draw [color=gray] (0,-0.22,1.25) node [] {$\xh_3$};
\draw [dashed,color=gray] (0,0,1) -- (0.5,0.5,0);			
\draw [thick,->] (1/3,1/3,1/3) -- (1/3+0.4,1/3+0.4,1/3+0.4);	
\draw [thick,->] (0,1/3,1/3) -- (-0.4,1/3,1/3);
\draw [thick,->] (1/3,0,1/3) -- (1/3,-0.4,1/3);
\draw [thick,->] (1/3,1/3,0) -- (1/3,1/3,-0.4);
\draw [color=gray] (1/3+0.25,1/3+0.4,1/3+0.25) node [] {\footnotesize $4$};		
\draw [color=gray] (-0.4+0.15,1/3+0.65,1/3+0.3) node [] {\footnotesize $3: \hat{f}_3, \hat{n}_3$};
\draw [color=gray] (1/3,-0.4-0.1,1/3) node [] {\footnotesize $2$};
\draw [color=gray] (1/3,1/3,-0.4-0.15) node [] {\footnotesize $1$};
\end{tikzpicture}
}
\mbox{
\begin{tikzpicture}[scale=1.3,tdplot_main_coords]
\draw (0,0.3,-1.15) node [] {\footnotesize 3D \nedelec};	
\draw (0,0.3,-1.5) node [] {\footnotesize (edges, \#dof=6)};
\draw [->] (0,0,0) -- (1.5,0,0);						
\draw [->] (0,0,0) -- (0,1.5,0);
\draw [->] (0,0,0) -- (0,0,1.4);
\draw (1,0,0) -- (0,1,0);							
\draw (0,1,0) -- (0,0,1);
\draw (0,0,1) -- (1,0,0);
\draw [fill,color=black] (0,0,0) circle(0.04);				
\draw [fill,color=black] (1,0,0) circle(0.04);
\draw [fill,color=black] (0,1,0) circle(0.04);
\draw [fill,color=black] (0,0,1) circle(0.04);
\draw [color=gray] (1.6,-0.1,0) node [] {$\xh_1$};			
\draw [color=gray] (0,1.7,0) node [] {$\xh_2$};
\draw [color=gray] (0,-0.22,1.25) node [] {$\xh_3$};
\draw [thick,->] (0.1,-0.1,0) -- (0.5,-0.1,0);				
\draw [thick,->] (-0.1,0.1,0.03) -- (-0.1,0.5,0.03);
\draw [thick,->] (-0.1,0.03,0.1) -- (-0.1,0.03,0.5);
\draw [thick,->] (0.8,0.4,-0.01) -- (0.5,0.7,-0.01);
\draw [thick,->] (0.5-0.1,-0.1,0.7-0.1) -- (0.8-0.1,-0.1,0.4-0.1);
\draw [thick,->] (0,0.8,0.4) -- (0,0.5,0.7);
\draw [color=gray] (0.1,-0.1,0.15) node [] {\footnotesize $1$};		
\draw [color=gray] (-0.1,0.3,0.2) node [] {\footnotesize $2$};
\draw [color=gray] (-0.1,0.18,0.5) node [] {\footnotesize $3$};
\draw [color=gray] (0.5,0.7,-0.15) node [] {\footnotesize $4$};
\draw [color=gray] (0.8-0.1,-0.23,0.4-0.1) node [] {\footnotesize $6$};
\draw [color=gray] (-0.35,0.9,0.75) node [] {\footnotesize $5: \hat{e}_5, \hat{t}_5$};
\end{tikzpicture}
}
\caption{Degrees of freedom of linear edge elements in the reference configuration $\hat{K}$.}
\label{fig:dofs}
\end{figure}
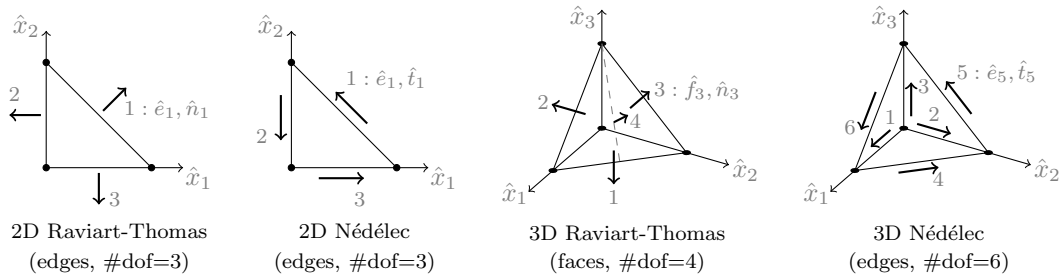

We denote by $\O$ an open, bounded, and connected Lipschitz domain in $\Rd$, where $d\in\{2,3\}$ denotes the space dimension. The divergence (2D and 3D) and rotation (3D) of a vector valued function $w: \O \rightarrow \Rd$ are defined as 
\begin{equation*}
	\dvg \, w := \sum_{i=1}^d \partial_i w_i
	\qquad \textrm{and} \qquad
	\curl \, w := \begin{pmatrix}
	\partial_2 w_3 - \partial_3 w_2 \\
	\partial_3 w_1 - \partial_1 w_3 \\
	\partial_1 w_2 - \partial_2 w_1 
	\end{pmatrix} .
\end{equation*}
We consider two types of rotation operators in 2D, the vector operator $\underline{\curl}$ and the scalar operator $\curl$
\begin{equation*}
	\underline{\curl} \, f := \begin{pmatrix} \partial_2 f \\ - \partial_1 f \end{pmatrix}
	\qquad \textrm{and} \qquad
	\curl \, w := \partial_1 w_2 - \partial_2 w_1
\end{equation*}
applied to a scalar function $f: \O \rightarrow \R$ and to a vector function $w: \O \rightarrow \R^{2}$.
The operator $\curlv$ is frequently called the ''co-gradient'' in literature, and is often denoted by $\nabla^\bot$. The operators give rise to the standard Sobolev spaces:
\begin{align*}
	\H(\dvg,\O) & := \{ v \in L^2(\O,\Rd) \mid \dvg \, v \in L^2(\O) \} , \\
	\H(\curl,\O) & := 
	\left\{
	\begin{array}{l c}
	\{ v \in L^2(\O,\R^3) \; \mid \; \curl \, v \in L^2(\O,\R^3) \} & \textrm{if} \; d=3 \\
	\{ v \in L^2(\O,\R^2) \; \mid \; \curl \, v \in L^2(\O) \}	& \textrm{if} \; d=2
	\end{array}
	\right. ,
\end{align*}
where $L^2$ denotes the space of square Lebesque integrable functions. We will denote the $\Ltwo$-norm of scalar and vector valued functions by $\Norms{\cdot} := \Norms{\cdot}_{\Ltwo(\O)}$. Assuming that $\Omega$ is discretized by a triangular (2D) or a tetrahedral (3D) mesh $\mathcal T$, Raviart-Thomas and \nedelec elements represent basis functions in $\H(\dvg,\mathcal T)$ and $\H(\curl,\mathcal T)$ spaces.

In the case of the lowest order (linear) Raviart-Thomas and \nedelec elements, there is one global degree of freedom (dof), i.e., one global basis function, related to either each edge (2D and 3D), or each face (3D) of a mesh $\mathcal{T}$. Due to construction, the global Raviart-Thomas basis functions and the \nedelec basis functions in 2D are nonzero only in the two elements who share the edge/face that is related to the basis function. In 3D the global \nedelec basis function is nonzero in all the elements sharing the related edge, and the number of these elements is usually more than two.

We denote the global edge/face basis functions by $\eta^\rt$ and $\eta^\ned$, and by $x = (x_1,x_2,x_3)^{\rm T}$ the spatial variable in $\O$. The notation for reference basis functions and spatial variable is obtained by simply adding the hat $\hat{\cdot}$, i.e., $\xh$ denotes the spatial variable in the reference element $\hat{K}$. We will use the unit triangle in 2D and the unit tetrahedron in 3D as the reference elements. We denote by $\hat{e}_i$ the $i$'th edge of the reference triangle or tetrahedron, and by $\hat{f}_i$ the $i$'th face of the reference tetrahedron. The numbering of the edges and faces, i.e., the numbering of the degrees of freedom in the reference elements, can be seen in Figure \ref{fig:dofs}. In the following, $F_K$ denotes the affine element mapping $F_K(\xh) := B_K\xh + b_K$ from the reference element $\hat{K}$ to an element $K$ in the mesh.

A finite element is defined by the triplet $\{\hat{K},\hat{R},\hat{A}\}$, where $\hat{K}$ is the reference configuration, $\hat{R}$ the finite space of functions defined on the reference configuration, and $\hat{A}$ is the set of linearly independent degrees of freedom. The reference configurations we have already chosen. We also need a mapping which takes functions from $\hat{R}$ and maps them from $\hat{K}$ to an element $K$ on the mesh $\mathcal T$. These mappings are called Piola mappings.

\subsection{Raviart-Thomas element}
The linear Raviart-Thomas element is based on the spaces (see, e.g., \cite{Ned,RT})
\begin{equation*}
	\textrm{\textbf{2D}:} \quad
	\hat{R} =
	\left\langle 
	\begin{pmatrix} 1 \\ 0 \end{pmatrix} ,
	\begin{pmatrix} 0 \\ 1 \end{pmatrix} ,
	\begin{pmatrix} \xh_1 \\ \xh_2 \end{pmatrix}
	\right\rangle ,
	\qquad
	\textrm{\textbf{3D}:} \quad
	\hat{R} =
	\left\langle 
	\begin{pmatrix} 1 \\ 0 \\ 0 \end{pmatrix} ,
	\begin{pmatrix} 0 \\ 1 \\ 0 \end{pmatrix} ,
	\begin{pmatrix} 0 \\ 0 \\ 1 \end{pmatrix} ,
	\begin{pmatrix} \xh_1 \\ \xh_2 \\ \xh_3 \end{pmatrix}
	\right\rangle ,
\end{equation*}
and the degrees of freedom for $\hat{u} \in \hat{R}$ read as
\begin{equation} \label{eq:RTdofs}
	\textrm{\textbf{2D}:} \quad
	\hat{A} = \left\{
	\hat{\alpha}_i(\hat{u}) = \int_{\hat{e}_i} \hat{n}_i \cdot \hat{u} \; \dsh \; , \quad i\in\{1,2,3\}
	\right\} ,
	\qquad
	\textrm{\textbf{3D}:} \quad
	\hat{A} = \left\{
	\hat{\alpha}_i(\hat{u}) = \int_{\hat{f}_i} \hat{n}_i \cdot \hat{u} \; \dsh \; , \quad i\in\{1,2,3,4\}
	\right\}
\end{equation}
for every edge $\hat{e}_i$ in 2D, or face $\hat{f}_i$ in 3D, in the corresponding reference elements $\hat{K}$. There are three dofs in 2D and four in 3D. Here $\hat{n}_i$ is the normal unit vector of the edge $\hat{e}_i$, or the face $\hat{f}_i$. Here one has to choose which of the two possible unit normal vectors to use. The standard choice of outer unit normals is depicted in Figure \ref{fig:dofs}. The requirement $\hat{\alpha}_i(\etah^\rt_j) = \delta_{ij}$ (where $\delta_{ij}$ is the Kronecker delta) gives us the reference basis functions of the Raviart-Thomas element:
\begin{align*}
	\textrm{\textbf{2D}:} \qquad &
	\etah^\rt_{1}(\xh) = \begin{pmatrix} \xh_1 \\ \xh_2 \end{pmatrix} , \quad 
	\etah^\rt_{2}(\xh) = \begin{pmatrix} \xh_1-1 \\ \xh_2 \end{pmatrix} , \quad 
	\etah^\rt_{3}(\xh) = \begin{pmatrix} \xh_1 \\ \xh_2-1 \end{pmatrix} ,
	\\
	\textrm{\textbf{3D}:} \qquad &
	\etah^\rt_{1}(\xh) = \begin{pmatrix} \xh_1 \\ \xh_2 \\ \xh_3-1 \end{pmatrix} , \quad 
	\etah^\rt_{2}(\xh) = \begin{pmatrix} \xh_1 \\ \xh_2-1 \\ \xh_3 \end{pmatrix} , \quad 
	\etah^\rt_{3}(\xh) = \begin{pmatrix} \xh_1-1 \\ \xh_2 \\ \xh_3 \end{pmatrix} , \quad
	\etah^\rt_{4}(\xh) = \begin{pmatrix} \xh_1 \\ \xh_2 \\ \xh_3 \end{pmatrix} .
\end{align*}
In order to preserve normal continuity of the reference basis functions, we need to use the so-called Piola mappings. The values and the divergence values are mapped as follows (see, e.g., \cite{BrFo}):
\begin{equation} \label{eq:RT0piola}
	\restr{\eta^\rt}{K}(x) = \frac{1}{\det B_K} B_K \; \etah^\rt( F_K^{-1}(x) )
	\qquad \textrm{and} \qquad
	\restr{\dvg \, \eta^\rt}{K}(x) = \frac{1}{\det B_K} \; \dvg \, \etah^\rt ( F_K^{-1}(x) ) .
\end{equation}


\subsection{\nedelec element}
The linear \nedelec element is based on the spaces (see, e.g., \cite{Ned,Schneebeli})
\begin{equation*}
	\textrm{\textbf{2D}:} \quad
	\hat{R} =
	\left\langle 
	\begin{pmatrix} 1 \\ 0 \end{pmatrix} ,
	\begin{pmatrix} 0 \\ 1 \end{pmatrix} ,
	\begin{pmatrix} \xh_2 \\ - \xh_1 \end{pmatrix}
	\right\rangle ,
	\qquad
	\textrm{\textbf{3D}:} \quad
	\hat{R} =
	\left\langle 
	\begin{pmatrix} 1 \\ 0 \\ 0 \end{pmatrix} ,
	\begin{pmatrix} 0 \\ 1 \\ 0 \end{pmatrix} ,
	\begin{pmatrix} 0 \\ 0 \\ 1 \end{pmatrix} ,
	\begin{pmatrix} 0 \\ \xh_3 \\ \xh_2 \end{pmatrix} ,
	\begin{pmatrix} \xh_3 \\ 0 \\ \xh_1 \end{pmatrix} ,
	\begin{pmatrix} \xh_2 \\ \xh_1 \\ 0 \end{pmatrix}
	\right\rangle ,
\end{equation*}
and the degrees of freedom for $\hat{u} \in \hat{R}$ in both dimensions are related to the edges of the elements:
\begin{equation} \label{eq:NEDdofs}
	\hat{A} = \left\{ \hat{\alpha}_i(\hat{u}) = \int_{\hat{e}_i} \hat{t}_i \cdot \hat{u} \; \dsh \; , \quad i\in\{1,2,\ldots\} \right\}
\end{equation}
for every edge $\hat{e}_i$ in the reference configuration $\hat{K}$. There are three dofs in 2D and six in 3D. Here $\hat{t}_i$ is the tangential unit vector of the edge $\hat{e}_i$. Similarly to the Raviart-Thomas element, one has to choose which direction for the unit tangential vectors to use. Our choice is depicted in Figure \ref{fig:dofs}. The requirement $\hat{\alpha}_i(\etah^\ned_j) = \delta_{ij}$ gives us the reference basis functions of the \nedelec element:
\begin{align*}
	\textrm{\textbf{2D}:} \qquad &
	\etah^\ned_{1}(\xh) = \begin{pmatrix} -\xh_2 \\ \xh_1 \end{pmatrix} , \quad 
	\etah^\ned_{2}(\xh) = \begin{pmatrix} -\xh_2 \\ \xh_1-1 \end{pmatrix} , \quad 
	\etah^\ned_{3}(\xh) = \begin{pmatrix} 1-\xh_2 \\ \xh_1 \end{pmatrix} ,
	\\
	\textrm{\textbf{3D}:} \qquad &
	\etah^\ned_{1}(\xh) = \begin{pmatrix} 1-\xh_3-\xh_2 \\ \xh_1 \\ \xh_1 \end{pmatrix} , \quad 
	\etah^\ned_{2}(\xh) = \begin{pmatrix} \xh_2 \\ 1-\xh_3-\xh_1 \\ \xh_2 \end{pmatrix} , \quad 
	\etah^\ned_{3}(\xh) = \begin{pmatrix} \xh_3 \\ \xh_3 \\ 1-\xh_2-\xh_1 \end{pmatrix} ,
	\\
	&
	\etah^\ned_{4}(\xh) = \begin{pmatrix} -\xh_2 \\ \xh_1 \\ 0 \end{pmatrix} , \quad
	\etah^\ned_{5}(\xh) = \begin{pmatrix} 0 \\ -\xh_3 \\ \xh_2 \end{pmatrix} , \quad
	\etah^\ned_{6}(\xh) = \begin{pmatrix} \xh_3 \\ 0 \\ -\xh_1 \end{pmatrix} .
\end{align*}
Again, we need to use a Piola mapping in order to preserve the tangential continuity (see, e.g., \cite{Ned,Schneebeli}). The values are mapped as follows:
\begin{equation} \label{eq:Ned0piola1}
	\restr{\eta^\ned}{K}(x) = B_K^{-\rm T} \; \etah^\ned ( F_K^{-1} (x) ) .
\end{equation}
The rotation is mapped differently depending on the dimension:
\begin{align}
	\textrm{\textbf{2D}:} & \qquad
	\restr{\curl \, \eta^\ned}{K}(x) = \frac{1}{\det B_K} \; \curl \, \etah^\ned ( F_K^{-1} (x) ) , \label{eq:Ned0piola2} \\
	\textrm{\textbf{3D}:} & \qquad
	\restr{\curl \, \eta^\ned}{K}(x) = \frac{1}{\det B_K} B_K \; \curl \, \etah^\ned ( F_K^{-1} (x) ) . \label{eq:Ned0piola3}
\end{align}


\subsection{Orientation of local degrees of freedom} \label{ssec:orientation}

In order to obtain the global basis functions $\eta^\rt$ and $\eta^\ned$, the transformations described in the previous sections are not enough. A global basis function is related to more than one element. No consideration has been yet made in making sure that the local orientation of the degrees of freedom \eqref{eq:RTdofs} and  \eqref{eq:NEDdofs} in these different elements is the same. The orientation must be same in order for the Raviart-Thomas and \nedelec elements to produce functions whose normal component, or tangential component (respectively) are continuous at element interfaces.

Take for example the Raviart-Thomas element in 2D. Let $K_n$ and $K_m$ be two elements in a mesh which share an edge (see Figure \ref{fig:signs}), and let $\eta^\rt$ be the global basis function related to this edge. We denote by $\etah^\rt_k$ and $\etah^\rt_l$ the reference basis functions which we will transform from $\hat{K}$ to $K_n$ and $K_m$ respectively, in order to obtain the global basis function.

By taking a look at the dofs \eqref{eq:RTdofs}, we see that we are always using the outer unit normals to compute the local basis functions. If we simply use the transformation \eqref{eq:RT0piola}, the normal component of the values at the edge might be the opposite of each other. This depends wether or not the element mappings $F_{K_n}$ and $F_{K_m}$ preserve orientation. If $\det B_{K_n} > 0$, and $\det B_{K_m} < 0$, the element mapping $F_{K_n}$ preserves the counter-clockwise orientation of the reference element, and $F_{K_m}$ is oriented clockwise. This means that on the common edge the orientation is in the same direction, and the transformation \eqref{eq:RT0piola} is enough for both elements. However, otherwise the orientation on the common edge will be the opposite, and one of the transformations must be multiplied by $-1$. The global basis function is thus obtained by
\begin{equation} \label{eq:RT0piola2}
	\restr{\eta^\rt}{K_n}(x) = \signs{K_n}{k} \frac{1}{\det B_{K_n}} B_{K_n} \; \etah^\rt_k( F_{K_n}^{-1}(x) )
	\qquad {\rm and} \qquad
	\restr{\eta^\rt}{K_m}(x) = \signs{K_m}{l} \frac{1}{\det B_{K_m}} B_{K_m} \; \etah^\rt_l( F_{K_m}^{-1}(x) ) ,
\end{equation}
where
\begin{align*}
	\signs{K_n}{k} = +1, \; \signs{K_m}{l} = +1 \qquad {\rm if} \qquad \det B_{K_n} > 0, \; \det B_{K_m} < 0 , \\
	\signs{K_n}{k} = +1, \; \signs{K_m}{l} = -1 \qquad {\rm if} \qquad \det B_{K_n} > 0, \; \det B_{K_m} > 0 , \\
	\signs{K_n}{k} = -1, \; \signs{K_m}{l} = +1 \qquad {\rm if} \qquad \det B_{K_n} < 0, \; \det B_{K_m} < 0 , \\
	\signs{K_n}{k} = -1, \; \signs{K_m}{l} = -1 \qquad {\rm if} \qquad \det B_{K_n} < 0, \; \det B_{K_m} > 0 .
\end{align*}
We call these values the sign data related to each of the two elements. Note that the above means that the global basis function is obtained simply by
\begin{equation*}
	\restr{\eta^\rt}{K_n}(x) = + \frac{1}{\abs{\det B_{K_n}}} B_{K_n} \; \etah^\rt_k( F_{K_n}^{-1}(x) )
	\qquad {\rm and} \qquad
	\restr{\eta^\rt}{K_m}(x) = - \frac{1}{\abs{\det B_{K_m}}} B_{K_m} \; \etah^\rt_l( F_{K_m}^{-1}(x) ) ,
\end{equation*}
but we will use \eqref{eq:RT0piola2} since it is more convenient to implement in program code.

The situation is the same for 3D Raviart-Thomas element and the 2D \nedelec element. For the \nedelec element in 3D one needs to be more careful since a global basis function may be nonzero in a relatively large patch of elements, and the relevant orientation is related to an edge. Note also that the same sign data must be used when transforming the divergence or rotation of the basis functions.


\subsection{Finite element matrices} \label{ssec:localassembly}
We are interested in assembly of the mass matrices $\mass^\rt, \mass^\ned$ and the stiffness matrices $\stiff^\rt, \stiff^\ned$ defined by
\begin{align*}
	\mass_{ij}^\rt & = \int_\Omega \eta^\rt_i \cdot \eta^\rt_j \, \dx , &
	\mass_{ij}^\ned & = \int_\Omega \eta^\ned_i \cdot \eta^\ned_j \, \dx , \\
	\stiff_{ij}^\rt & = \int_\Omega \dvg \, \eta^\rt_i \; \dvg \, \eta^\rt_j \, \dx , &
	\stiff_{ij}^\ned & = \int_\Omega \curl \, \eta^\ned_i \cdot \curl\, \eta^\ned_j \, \dx ,
\end{align*}
where the indexes $i$ and $j$ are the global numbering of the degrees of freedom, i.e., they are related to the edges or faces of a mesh. By using the Piola mappings (with correct orientations), we are able to assemble the local matrices using the reference element.

By using \eqref{eq:RT0piola2}, the local matrices related to the global matrices $\mass^\rt$ and $\stiff^\rt$ can be calculated on each element $K \in \mathcal T$ by
\begin{align}
	\mass_{kl}^{\rt,K} & = \frac{1}{\abs{\det B_K}} \int_{\hat{K}} \sign{k} B_K \etah_k^\rt(\xh) \cdot \sign{l} B_K \etah^\rt_l(\xh) \; \dxh , \nonumber 
	\\
	\stiff_{kl}^{\rt,K} & = \frac{1}{\abs{\det B_K}} \int_{\hat{K}} \sign{k} \dvg \, \etah_k^\rt(\xh) \;\; \sign{l} \dvg \, \etah^\rt_l(\xh) \; \dxh , \label{eq:RT0local2}
\end{align}
where $\hat{K}$ is the reference element. The indexes $k$ and $l$ run through all the local basis functions in the element: $k,l \in \{1,2,3\}$ in 2D, and $k,l \in \{1,2,3,4\}$ in 3D.

Similarly, by using \eqref{eq:Ned0piola1} (and considering the correct orientations, see Section \ref{ssec:orientation}) the local mass matrices related to the global mass matrix $\mass^\ned$ can be calculated on each element $K \in \mathcal T$ by
\begin{equation*}
	\mass_{kl}^{\ned,K} = \abs{\det B_K} \int_{\hat{K}} \sign{k} B_K^{-\rm T} \etah_k^\ned(\xh) \cdot \sign{l} B_K^{-\rm T} \etah^\ned_l(\xh) \; \dxh ,
\end{equation*}
and by using \eqref{eq:Ned0piola2}--\eqref{eq:Ned0piola3} the local stiffness matrices related to the global stiffness matrix $\stiff^\ned$ can be calculated by
\begin{align*}
	\textrm{\textbf{2D}:} & \qquad
	\stiff_{kl}^{\ned,K} = \frac{1}{\abs{\det B_K}} \int_{\hat{K}} \sign{k} \curl \, \etah_k^\ned(\xh) \;\; \sign{l} \curl \, \etah^\ned_l(\xh) \; \dxh , \\
	\textrm{\textbf{3D}:} & \qquad
	\stiff_{kl}^{\ned,K} = \frac{1}{\abs{\det B_K}} \int_{\hat{K}} \sign{k} B_K \curl \, \etah_k^\ned(\xh) \cdot \sign{k} B_K \curl \, \etah^\ned_l(\xh) \; \dxh .
\end{align*}
The indexes $k$ and $l$ run through all the local basis functions in the reference element: $k,l \in \{1,2,3\}$ in 2D, and $k,l \in \{1,2,3,4,5,6\}$ in 3D.


\section{Implementation of edge elements}
\begin{figure}
\centering
\mbox{
\begin{tikzpicture}[scale=1.4]
\draw [fill,color=white] (2.8,0) circle(0);      
\draw [line width=0.5mm] (0,0) -- (1,0);	
\draw [line width=0.5mm] (0,0) -- (0,1);
\draw [line width=0.5mm] (1,0) -- (0,1);
\draw [color=gray] (1,0) -- (1,1);
\draw [color=gray] (1,1) -- (0,1);
\draw [color=gray] (0,0) -- (-1,0); %
\draw [color=gray] (0,1) -- (-1,0);
\draw [color=gray] (-1,0) -- (-1,1); %
\draw [color=gray] (0,1) -- (-1,1);
\draw [color=gray] (1,0) -- (2,0); %
\draw [color=gray] (1,0) -- (2,1);
\draw [color=gray] (2,0) -- (2,1);
\draw [color=gray] (2,1) -- (1,1);
\draw [color=gray] (-1,1) -- (-1,2); %
\draw [color=gray] (0,1) -- (0,2);
\draw [color=gray] (1,1) -- (1,2);
\draw [color=gray] (2,1) -- (2,2);
\draw [color=gray] (0,2) -- (-1,2); %
\draw [color=gray] (1,2) -- (0,2);
\draw [color=gray] (2,2) -- (0,2);
\draw [color=gray] (-1,1) -- (0,2); %
\draw [color=gray] (0,1) -- (1,2);
\draw [color=gray] (1,1) -- (2,2);
\draw [color=gray] (-1,0) -- (-1,-1); %
\draw [color=gray] (0,0) -- (0,-1);
\draw [color=gray] (1,0) -- (1,-1);
\draw [color=gray] (2,0) -- (2,-1);
\draw [color=gray] (-1,-1) -- (0,-1); %
\draw [color=gray] (0,-1) -- (1,-1);
\draw [color=gray] (2,-1) -- (1,-1);
\draw [color=gray] (-1,-1) -- (0,0); %
\draw [color=gray] (0,-1) -- (1,0); %
\draw [color=gray] (1,-1) -- (2,0); %

\draw [fill,color=black] (0,0) circle(0.03);			
\draw [fill,color=black] (1,0) circle(0.03);
\draw [fill,color=black] (0,1) circle(0.03);
\draw [color=black] (0.3,0.3) node [] {\footnotesize $K_n$};
\draw [color=black] (1.2,-0.2) node [] {\small $\mathring{b}$};
\draw [color=black] (-0.17,1.2) node [] {\small $\mathring{c}$};
\draw [color=black] (-0.17,-0.2) node [] {\small $\mathring{a}$};
\draw [color=black] (-0.17,0.5) node [] {\small $\up{b}$};
\draw [color=black] (0.5,-0.2) node [] {\small $\up{c}$};
\draw [color=black] (0.64,0.64) node [] {\small $\up{a}$};
\end{tikzpicture}
}
\mbox{
\begin{tikzpicture}[scale=1.34]
\draw [color=black] (4.3,1.5) node [] {\footnotesize
	\url{elems2nodes} =
	$\begin{bmatrix}
        \color{gray}\vdotss{1} & \color{gray}\vdotss{1} & \color{gray}\vdotss{1} \\
        \mathring{a} & \mathring{b} & \mathring{c} \\
        \color{gray}\vdotss{1} & \color{gray}\vdotss{1} & \color{gray}\vdotss{1}
	\end{bmatrix}
	\begin{matrix}
        \color{gray}\vdotss{1}\\
        (\textrm{row} \; n) \\
        \color{gray}\vdotss{1}
	\end{matrix}$
};
\draw [color=black] (8.3,1.5) node [] {\footnotesize
	\url{elems2edges} =
	$\begin{bmatrix}
        \color{gray} \vdotss{1} & \color{gray} \vdotss{1} & \color{gray} \vdotss{1} \\
        \up{a} & \up{b} & \up{c} \\
        \color{gray} \vdotss{1} & \color{gray} \vdotss{1} & \color{gray} \vdotss{1}
	\end{bmatrix}
	\begin{matrix}
        \color{gray} \vdotss{1} \\
        (\textrm{row} \; n) \\
        \color{gray} \vdotss{1}
	\end{matrix}$
};
\draw [color=black] (4.24,-0.1) node [] {\footnotesize
	\url{nodes2coord} =
	$\begin{bmatrix}
        \color{gray}\vdotss{1} & \color{gray}\vdotss{1} \\
        \mathring{a}_{1} & \mathring{a}_{2} \\
        \color{gray}\vdotss{1} & \color{gray}\vdotss{1} \\
	\mathring{b}_{1} & \mathring{b}_{2} \\
        \color{gray}\vdotss{1} & \color{gray}\vdotss{1} \\
	\mathring{c}_{1} & \mathring{c}_{2} \\
        \color{gray}\vdotss{1} & \color{gray}\vdotss{1}
        \end{bmatrix}
	\begin{matrix}
        \color{gray}\vdotss{1} \\
        (\textrm{row} \; \mathring{a}) \\
        \color{gray}\vdotss{1} \\
	(\textrm{row} \; \mathring{b}) \\
        \color{gray}\vdotss{1} \\
	(\textrm{row} \; \mathring{c}) \\
        \color{gray}\vdotss{1}
	\end{matrix}$
};
\draw [color=black] (8.24,-0.1) node [] {\footnotesize
	\url{edges2nodes} =
	$\begin{bmatrix}
        \color{gray}\vdotss{1} & \color{gray}\vdotss{1} \\
        \up{a}_{1} & \up{a}_{2} \\
        \color{gray}\vdotss{1} & \color{gray}\vdotss{1} \\
	\up{b}_{1} & \up{b}_{2} \\
        \color{gray}\vdotss{1} & \color{gray}\vdotss{1} \\
	\up{c}_{1} & \up{c}_{2} \\
        \color{gray}\vdotss{1} & \color{gray}\vdotss{1}
        \end{bmatrix}
	\begin{matrix}
        \color{gray}\vdotss{1} \\
        (\textrm{row} \; \up{a}) \\
        \color{gray}\vdotss{1} \\
	(\textrm{row} \; \up{b}) \\
        \color{gray}\vdotss{1} \\
	(\textrm{row} \; \up{c}) \\
        \color{gray}\vdotss{1}
	\end{matrix}$
};
\draw [color=gray] (7.27,-0.85) node [] {\scriptsize
	\url{nodes2coord}
};
\draw [<-] (3.3,0.2) -- (3.3,1.2);
\draw [<-] (7.3,0.2) -- (7.3,1.2);
\draw [color=gray] [<-] (7.3,-0.6) -- (7.3,-0.35);
\end{tikzpicture}
}
\caption{Elements by their nodes and edges, i.e., global numbering of degrees of freedom for 2D linear finite elements.}
\label{fig:dofs2}
\end{figure}
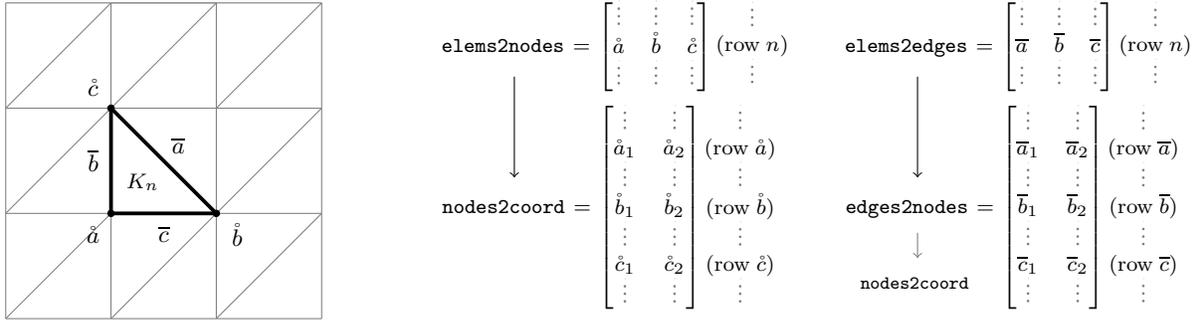

We denote by $\# \omega$ the number of elements in the set $\omega$, and by $\nodes,\edges,\faces$, and $\elems$ the sets of nodes, edges, faces, and elements, respectively.  Note that faces $\faces$ exist only in 3D. We need the following structures representing the mesh in order to implement edge elements. The second column states the size of the structure, and the third column the meaning of the structure.
\\ \\
\begin{tabular}{l|l|l}
\verb+nodes2coord+ & $\#\nodes \times 2/3$ & nodes defined by their two/three coordinates in 2D/3D (in \cite{RaVa} \verb+coordinates+) \\
\verb+edges2nodes+ & $\#\edges \times 2$ & edges defined by their two nodes in 2D/3D \\
\verb+faces2nodes+ & $\#\faces \times 3$ & faces defined by their three nodes in 3D
\end{tabular}
\vskip+1em
\noindent
With these matrices available, we can then express every element by the list of its nodes, edges, or faces:\\
\\
\begin{tabular}{l|l|l}
\verb+elems2nodes+ & $\#\elems \times 3/4$ & elements by their three/four nodes in 2D/3D (in \cite{RaVa} \verb+elements+) \\
\verb+elems2edges+ & $\#\elems \times 3/6$ & elements by their three/six edges in 2D/3D \\
\verb+elems2faces+ & $\#\elems \times 4$ & elements by their four faces in 3D
\end{tabular}
\vskip+1em
In 2D both the linear Raviart-Thomas element and the linear \nedelec element have a degree of freedom related to each of the three edges of the reference triangle, totalling three dofs. In 3D the linear \nedelec element has a dof related to each of the six edges, and the Raviart-Thomas element will have a dof related to each of the four faces. Thus, the global numbering of degrees of freedom is given by the row indices of \verb+edges2nodes+ or \verb+faces2nodes+. For a particular element $K$ in the mesh, the global dofs related to it are then given by the structures \verb+elems2edges+ or \verb+elems2faces+, respectively. For nodal elements the global numbering of dofs is given by the row indices of \verb+nodes2coord+, and the dofs related to a particular element $K$ are given by \verb+elems2nodes+. In Figure \ref{fig:dofs2} we have further illustrated the structure of the mesh data in 2D.

\begin{figure} [h]
\centering
\mbox{
\begin{tikzpicture}[scale=1.2]
\draw [fill,color=white] (2,0) circle(0);      
\draw (0.15,-0.6) node [] {\footnotesize 2D \nedelec};
\draw (0.15,-0.9) node [] {\footnotesize (tangential direction)};
\draw (0,0) -- (0,1.2); 
\draw (-1,0.6) -- (0,0);
\draw (-1,0.6) -- (0,1.2);
\draw (0.3,0) -- (0.3,1.2); 
\draw (1.3,0.6) -- (0.3,0);
\draw (1.3,0.6) -- (0.3,1.2);
\draw [line width=0.5mm,->] (0.15,1) -- (0.15,0.2);	
\draw [<-] (-0.15,0.9) -- (-0.15,0.3);	
\draw [->] (0.45,0.9) -- (0.45,0.3);	
\draw [color=gray] (-0.4,0.6) node [] {$-$};		
\draw [color=gray] (0.7,0.6) node [] {$+$};		
\end{tikzpicture}
}
\mbox{
\begin{tikzpicture}[scale=1.2]
\draw [fill,color=white] (2.1,0) circle(0);      
\draw (0.3,-0.6) node [] {\footnotesize 2D Raviart-Thomas};
\draw (0.3,-0.9) node [] {\footnotesize (normal direction)};
\draw (0,0) -- (0,1.2); 
\draw (-1,0.6) -- (0,0);
\draw (-1,0.6) -- (0,1.2);
\draw (0.6,0) -- (0.6,1.2); 
\draw (1.6,0.6) -- (0.6,0);
\draw (1.6,0.6) -- (0.6,1.2);
\draw [line width=0.5mm,<-] (0.1,0.6) -- (0.5,0.6);	
\draw [->] (0,0.4) -- (0.4,0.4);	
\draw [<-] (0.2,0.8) -- (0.6,0.8);	
\draw [color=gray] (0.2,0.2) node [] {$-$};		
\draw [color=gray] (0.4,1.0) node [] {$+$};		
\draw [color=black] (1.15,0.6) node [] {\footnotesize $K_n$};
\draw [color=black] (-0.5,0.6) node [] {\footnotesize $K_m$};
\draw [color=black] (0.75,0.6) node [] {\small $\up{b}$};
\draw [color=black] (1.2,1.05) node [] {\small $\up{a}$};
\draw [color=black] (1.2,0.15) node [] {\small $\up{c}$};
\draw [fill,color=black] (0.6,1.2) circle(0.03);	
\draw [fill,color=black] (0.6,0) circle(0.03);
\draw [fill,color=black] (1.6,0.6) circle(0.03);
\draw [color=black] (0.5,1.4) node [] {\small $\mathring{c}$};
\draw [color=black] (0.5,-0.2) node [] {\small $\mathring{a}$};
\draw [color=black] (1.76,0.6) node [] {\small $\mathring{b}$};
\end{tikzpicture}
}
\mbox{
\begin{tikzpicture}[scale=1.25]
\draw [color=black] (0,0) node [] {\footnotesize
	\url{elems2edges} =
	$\begin{bmatrix}
        \color{gray} \vdotss{1} & \color{gray} \vdotss{1} & \color{gray} \vdotss{1} \\
        \up{a} & \up{b} & \up{c} \\
        \color{gray} \vdotss{1} & \color{gray} \vdotss{1} & \color{gray} \vdotss{1} \\
	\end{bmatrix}
	\begin{matrix}
        \color{gray} \vdotss{1} \\
        (\textrm{row} \; n) \\
        \color{gray} \vdotss{1}
	\end{matrix}$
};
\draw [color=black] (0.52,-1.1) node [] {\footnotesize
	\url{signs_e} =
	$\begin{bmatrix}
        \color{gray} \vdotss{1} & \color{gray} \vdotss{1} & \color{gray} \vdotss{1} \\
        \up{a}_{\pm} & +1 & \up{c}_{\pm} \\
        \color{gray} \vdotss{1} & \color{gray} \vdotss{1} & \color{gray} \vdotss{1} \\
	\end{bmatrix}
	\begin{matrix}
        \color{gray} \vdotss{1} \\
        (\textrm{row} \; n) \\
        \color{gray} \vdotss{1}
	\end{matrix}$
};
\draw [->] (-0.9,-0.3) -- (-0.9,-0.8);
\end{tikzpicture}
}
\caption{Orientation of 2D edge elements sharing an edge, when both elements are oriented counter-clockwise, and $\mathring{c} > \mathring{a}$. The thick line denotes the "positive direction".}
\label{fig:signs}
\end{figure}
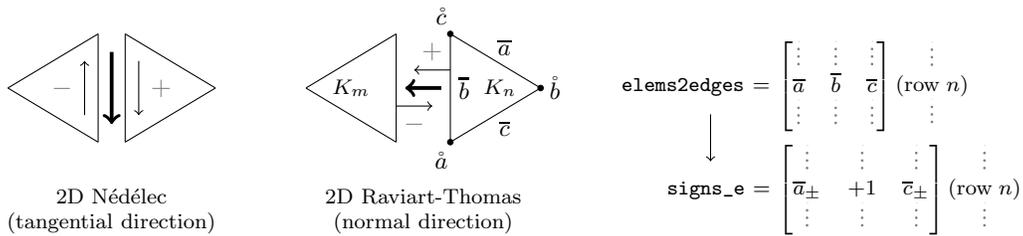

Since the degrees of freedom are integrals over edges or faces, we need to pay attention to orientation (see Sections \ref{ssec:orientation} and \ref{ssec:localassembly}). In practice we need to know how every edge/face of every element is oriented. Orientation is naturally given either by $+1$ or $-1$. We need the following structures:\\
\\
\begin{tabular}{l|l|l}
\verb+signs_e+ & $\#\elems \times 3/6$ & $+1$ or $-1$ for every edge of an element, corresponding to \verb+elems2edges+ \\
\verb+signs_f+ & $\#\elems \times 4$ & $+1$ or $-1$ for every face of an element in 3D, corresponding to \verb+elems2faces+
\end{tabular}
\vskip+1em
In 2D obtaining the sign data for an element $K_i$ can be conveniently done by examining \verb+elems2nodes(i,:)+. The first edge of $K_i$ (\verb+elems2edges(i,1)+) is the edge from node 2 (\verb+elems2nodes(i,2)+) to node 3 (\verb+elems2nodes(i,3)+). We can then simply agree that if the global node indices satisfy \verb+elems2nodes(i,2)+ $>$ \verb+elems2nodes(i,3)+, we assign \verb+signs_e(i,1)+ $=1$, and \verb+signs_e(i,1)+ $=-1$ otherwise. This gives us the signs, or their opposites, as described in Section \ref{ssec:orientation}. This sign data can be used for both Raviart-Thomas element and the \nedelec element in 2D. The data structures are illustrated in Figure \ref{fig:signs}.

The procedure of determinning the signs for the 3D elements is straightforwads as well, but we will not comment on it here. In our software package the edges in 2D and 3D are calculated by the function \verb+get_edges()+ and the orientation related to edges is calculated by the function \verb+signs_edges()+. In 3D the faces are calculated by the function \verb+get_faces()+, and the orientation related to faces is calculated by \verb+signs_faces()+.

\subsection{Vectorized integration procedure}

As stated in the introduction, the key idea of this paper is to vectorize the integration procedure of an arbitrary function on an arbitrary mesh. The main ingredient is how to efficiently use integration quadratures via the reference element. We demonstrate our idea by explaining how to calculate the (squared) $\Ltwo$-norm $\left\Vert \verb+f+ \right\Vert^2$ of a function \verb+f+ $\in \Ltwo(\O)$. Provided that we have the structures \verb+nodes2coord+ and \verb+elems2nodes+ available, this is achieved (in the folder \verb+/example_majorant/+) with the following two lines:
\begin{verbatim}
       [B_K,b_K,B_K_det] = affine_transformations(nodes2coord,elems2nodes);
       f_L2norm          = norm_L2( elems2nodes, B_K, b_K, B_K_det, f );
\end{verbatim}
On the first line we obtain the affine transformations $F_K(\xh) = B_K\xh + b_K$ and determinants of $B_K$ for all elements $K$:\\
\\
\begin{tabular}{l|l|l}
\verb+B_K+ & $d \times d \times\#\elems$ & matrix parts $B_K$ \\
\verb+b_K+ & $\#\elems \times d$ & vector parts $b_K$ \\
\verb+B_K_det+ & $\#\elems \times 1$ & the determinants $\det B_K$
\end{tabular}
\vskip+1em
\noindent
The calculation of this data is done in a vectorized manner. On the second line we calculate the norm. The code of the function \verb+/example_majorant/norm_L2.m+ is
\begin{verbatim}
       function fnorm = norm_L2( elems2nodes, B_K, b_K, B_K_det, f );
    1  B_K_detA   = abs(B_K_det);
    2  dim        = size(B_K,1);
    3  [ip,w,nip] = intquad(6,dim);
    4  nelems     = size(elems2nodes,1);
    5  fnorm      = zeros(nelems,1);
    6  for i=1:nip
    7      F_K_ip = squeeze(amsv(B_K, ip(i,:)))' + b_K;
    8      fval   = f(F_K_ip);
    9      fnorm  = fnorm + w(i) .* B_K_detA .* fval.^2;
   10  end
   11  fnorm = sum(fnorm);
\end{verbatim}
On line 2 we deduce the dimension of the mesh. On line 3 the function \verb+[ip,w,nip] = intquad(po,dim)+ returns an integration quadrature of order \verb+po+ in the reference element. We use integration quadratures for triangles and tetrahedrons from \cite{Du} and \cite{ZhCuLi}, respectively. In this example we use quadrature order 6, so the calculation of the $\Ltwo$-norm is exact (up to machine precision) for polynomials of order 3 and less. The quadrature consists of the integration points \verb+ip+ and the weighs \verb+w+. The variable \verb+nip+ is the number of integration points. On the lines 4 and 5 we deduce the number of elements in the mesh in order to initialize the structure \verb+fnorm+.

Note that the \verb+for+-loop on line 6 is not over elements, but over integration points. This is what we mean by vectorization of the \verb+for+-loop over elements. Essentially we are replacing this loop with another, much smaller loop. Of course, since all the affine mappings and other data has to be available for all elements at the same time, this method requires more system memory.

On line 7 we transform the \verb+i+'th integration point to the mesh for all elements $K$ at the same time, and put this data into the  structure \verb+F_K_ip+. We have used here some functionality from the folder \verb+/path/library_vectorization/+, which was also used in the vectorization of nodal elements in \cite{RaVa}. This folder contains functions which perform certain operations between matrices and vectors, and does them in a vectorized manner. The function \verb+amsv.m+ from this folder takes in the matrices \verb+B_K+ and does the necessary multiplication with the \verb+i+'th integration point \verb+ip(i,:)+ for all entries simultaneously. On line 8 we calculate the values of the function \verb+f+ on all of these points, and on line 9 we add the contributions of the \verb+i+'th integration point to \verb+fnorm+.

After going through all the integration points, the structure \verb+fnorm+ contains the elementwise contributions of the norm, i.e., \verb+fnorm(i)+$ = \left\Vert \verb+f+ \right\Vert^2_{K_i}$, where $K_i$ is the element described by its nodes in \verb+elems2nodes(i,:)+. On the last line the elementwise contributions are summed together to obtain $\left\Vert \verb+f+ \right\Vert^2$.


\subsection{Vectorized finite element assembly routine}

The vectorized integration procedure of the previous section can be directly applied for finite element matrix assembly routines. As an example, we go through the needed program code for calculating the stiffness matrix $\stiff^\rt$ with Raviart-Thomas elements. We assume we have the mesh in the form of the structures \verb+nodes2coord+ and \verb+elems2nodes+, i.e., we have the node coordinates, and the representation of elements by their nodes.
\begin{verbatim}
       [B_K,~,B_K_det]           = affine_transformations(nodes2coord,elems2nodes);
       [elems2faces,faces2nodes] = get_faces(elems2nodes);
       signs_f                   = signs_faces(nodes2coord,elems2faces,faces2nodes,B_K);
\end{verbatim}
On the first line we obtain the affine transformation matrices and the determinants. On the second line we obtain the structure \verb+elems2faces+, which is the representation of elements by their faces. Note that indeed the numbers in \verb+elems2faces+ are indices to \verb+faces2nodes+. More importantly, \verb+elems2faces+ is the global numbering of the degrees of freedom for all elements $K$ in the mesh. On the third line we calculate the orientations for faces in 3D. Then, we call
\begin{verbatim}
       K_RT0 = stiffness_matrix_RT0(elems2faces,B_K_det,signs_f);
       M_RT0 = mass_matrix_RT0(elems2faces,B_K,B_K_det,signs_f);
\end{verbatim}
to assemble the stiffness and mass matrices. The main part of the function \verb+stiffness_matrix_RT0+ is the vectorized assembly routine:
\begin{verbatim}
       function STIFF = stiffness_matrix_RT0( elems, B_K_det, signs )
    1  dim             = size(elems,2)-1;
    2  nelems          = size(elems,1);
    3  B_K_detA        = abs(B_K_det);
    4  [ip,w,nip]      = intquad(1,dim);
    5  [~,dval,nbasis] = basis_RT0(ip);
    6  STIFF           = zeros(nbasis,nbasis,nelems);
    7  for i=1:nip
    8      for m=1:nbasis
    9          for k=m:nbasis
   10              STIFF(m,k,:) = squeeze(STIFF(m,k,:)) + ...
   11                             w(i) .* B_K_detA.^(-1) .* ...
   12                             ( signs(:,m) .* dval(i,:,m) ) .* ...
   13                             ( signs(:,k) .* dval(i,:,k) );
   14          end
   15      end
   16  end
   17  STIFF = copy_triu(STIFF);
   18  ...
\end{verbatim}
Note that this function does the assembly in both 2D and 3D, depending on the input variable \verb+elems+.

On lines 1--4 we deduce the dimension of the problem, deduce the number of elements in the mesh, calculate the absolute values of the determinants, and obtain the first order integration quadrature on the reference element. This is enough since for the linear Raviart-Thomas element the basis function divergences are constants. The function \verb+[val,dval,nbasis] = basis_RT0(ip)+ returns the values \verb+val+ and divergence values $\verb+dval+$ of the linear Raviart-Thomas reference basis functions at the integration points. Since we are assembling the stiffness matrix, we need only the divergence values. The variable \verb+nbasis+ is the number of basis functions per element. On line 6, the variable \verb+STIFF+ is initialized to be of suitable size to contain all the local element matrices.

Note again that the outer \verb+for+-loop on line 7 is not over elements, but over integration points. On lines 10--13 we assemble the local matrix entry \verb+(m,k)+ (for the integration point \verb+i+) for all elements at the same time. The assembly is done according to \eqref{eq:RT0local2}. Note that since the matrix is symmetric, it is sufficient to assemble only the diagonal and upper triangular entries, hence the indexing on the loop in line 9 begins from the previous loop index \verb+m+, and not 1. On line 17 the symmetric entries are copied to the lower triangular part of \verb+STIFF+. After this, the global matrix is assembled from the local matrices in \verb+STIFF+, but this part of the code we have excluded here.

This assembly routine consists only of the normal matrix operations of MATLAB. However, on most of the assembly routines we need to perform more complicated array operations. This functionality is provided by functions in \verb+/path/library_vectorization/+.

\subsection{Performance in 2D and 3D}
For investigating the performance of our vectorized assembly routines, we chose an L-shaped domain in 2D, and the unit cube for 3D. The results were performed with MATLAB 7.13.0.564 (R2011b) on a computer with 64 Intel(R) Xeon(R) CPU E7-8837 processors running at 2.67GHz, and 1 TB system memory. The computer is located at the University of Jyv\"askyl\"a. Results can be seen in Tables \ref{tab:2D_runtimes} and \ref{tab:3D_runtimes}.

Uniform refinement results in 4 times more triangles in 2D, and 8 times more tetrahedra in 3D. Thus, in each refinement step the optimal increase in time would be 4 in 2D and 8 in 3D. We see from Tables \ref{tab:2D_runtimes} and \ref{tab:3D_runtimes} that both 2D and 3D assembly routines scale with satisfactory performance as the problem size is increased. In 2D, on level 14 we already had over 2.4 billion elements, and the 1 TB system memory was still occupied by level 13 matrices. This forced the computer to start using swap memory, which considerably slowed the calculation of the new $\sim 2.4\;\textrm{billion} \times 2.4\;\textrm{billion}$ matrices for level 14.

It is also notable that in 3D the calculation of the matrices $\stiff^\ned$ and $\mass^\ned$ for the \nedelec element  takes over twice the time compared to the calculation of $\stiff^\rt$ and $\mass^\rt$ even though there are more degrees of freedom for the Raviart-Thomas matrices. The reason becomes evident when comparing the amount of algebraic operations that need to be calculated: for example, the divergences of Raviart-Thomas basis functions in 3D are scalar valued, but the rotations of \nedelec basis functions in 3D are vector valued.

\begin{table}[h]
\center
{\small
\begin{tabular}{l||r|r r r r r r r r}
   & size of & \multicolumn{8}{c}{assembly of} \\
level & matrices & \multicolumn{2}{c}{$\stiff^\rt$}  & \multicolumn{2}{c}{$\mass^\rt$} & \multicolumn{2}{c}{$\stiff^\ned$} & \multicolumn{2}{c}{$\mass^\ned$} \\
\hline
5 & 9 344 & 0.03  &  -  &  0.06  &  -  &  0.03  & -  &  0.03  &  -  \\
6 & 37 120 & 0.11  &  (3.6)  &  0.51  &  (8.5)  &  0.11  &  (3.6)  &  0.47  &  (15.6)  \\
7 & 147 968 & 0.41  &  (3.7)  &  1.08  &  (2.1)  &  0.40  &  (3.6)  &  1.02  &  (2.1)  \\
8 & 590 848 & 1.70  &  (4.1)  &  3.59  &  (3.3)  &  1.82  &  (4.5)  &  3.65  &  (3.5)  \\
9 & \textbf{2} 361 344 & 7.49  &  (4.4)  &  12.82  &  (3.5)  &  7.49  &  (4.1)  &  12.94  &  (3.5)  \\
10 & \textbf{9} 441 280 & 30.89  &  (4.1)  &  52.09  &  (4.0)  &  30.83  &  (4.1)  &  54.86  &  (4.2)  \\
11 & \textbf{37} 756 928 & 132.95  &  (4.3)  &  216.64  &  (4.1)  &  132.56  &  (4.2)  &  230.44  &  (4.2)  \\
12 & \textbf{151} 011 328 & 597.37  &  (4.4)  &  919.36  &  (4.2)  &  583.86  &  (4.4)  &  931.79  &  (4.0)  \\
13 & \textbf{604} 012 544 & 2620.11  &  (4.3)  &  3969.16  &  (4.3)  &  2840.51  &  (4.8)  &  4121.33  &  (4.4)  \\
14 & \textbf{2 415} 984 640 & 18333.25  &  (6.9)  &  33328.58  &  (8.3)  &  26781.41  &  (9.4)  &  37009.85  &  (8.9)
\end{tabular}
\caption{2D assembly times (in seconds) for an L-shaped domain $\Omega := (0,1)^2 \setminus (1/2,1)^2$. Values in brackets are the increase in time compared to the previous step (the optimal increase is 4).}
\label{tab:2D_runtimes}
}
\end{table}

\begin{table}[h]
\center
{\small
\begin{tabular}{l||r|r r r r ||r| r r r r}
   & size of & \multicolumn{4}{c||}{assembly of} & size of & \multicolumn{4}{c}{assembly of} \\
level & matrices & \multicolumn{2}{c}{$\stiff^\rt$}  & \multicolumn{2}{c||}{$\mass^\rt$} & matrices & \multicolumn{2}{c}{$\stiff^\ned$} & \multicolumn{2}{c}{$\mass^\ned$} \\
\hline
1  &  2 808  &  0.02  &  -  &  0.09  &  -  &  1 854  &  0.05  &  -  &  0.09  &  -  \\
2  &  21 600  &  0.14  &  (7.0)  &  0.39  &  (4.3)  &  13 428  &  0.30  &  (6.0)  &  0.79  &  (8.7)  \\
3  &  169 344  &  0.82  &  (5.8)  &  2.18  &  (5.5)  &  102 024  &  1.92  &  (6.4)  &  4.53  &  (5.7)  \\
4  &  \textbf{1} 340 928  &  7.15  &  (8.7)  &  15.35  &  (7.0)  &  795 024  &  15.44  &  (8.0)  &  33.43  &  (7.3)  \\
5  &  \textbf{10} 672 128  &  59.37  &  (8.3)  &  125.71  &  (8.1)  &  \textbf{6} 276 384  &  129.91  &  (8.4)  &  282.14  &  (8.4)  \\
6  &  \textbf{85} 155 840  &  503.89  &  (8.4)  &  1054.49  &  (8.3)  &  \textbf{49} 877 568  &  1125.08  &  (8.6)  &  2291.50  &  (8.1)  \\
7  &  \textbf{680} 361 984  &  4437.84  &  (8.8)  &  8717.70  &  (8.2)  &  \textbf{397} 689 984  &  10232.01  &  (9.0)  &  20028.06  &  (8.7)
\end{tabular}
\caption{3D assembly times (in seconds) for the unit cube $\Omega := (0,1)^3$. Values in brackets are the increase in time compared to the previous step (the optimal increase is 8).}
\label{tab:3D_runtimes}
}
\end{table}


\section{Examples of vectorized FEM computations using edge elements}


\subsection{Minimization of functional majorant using Raviart-Thomas elements}

Let us consider a scalar boundary value (Poisson's) problem
\begin{align*}
	- \triangle u & = f && \mbox{in } \O, \\
	\qquad u &= 0 &&  \mbox{on } \partial \O ,
\end{align*}
for a function $u \in \Hoz := \{ v \in \Ltwo(\O) \mid \nabla v \in \Ltwo(\O,\Rd), \; v = 0 \; \textrm{on} \; \partial\Omega \}$ and a given right hand side $f \in \Ltwo(\O)$. The exact solution $u$ is sought from the weak formulation 
\begin{equation} \label{eq:laplacegeneral}
	\int_\O \nabla u \cdot \nabla w \, \dx = \int_\O f w \, \dx \qquad \forall w \in \Hoz .
\end{equation}
Assume that $v \in \Hoz$ is an approximation of the exact solution $u$ of \eqref{eq:laplacegeneral}. Then, the functional type a posteriori error estimate from \cite{Re2004} states that
\begin{equation}
  \Norms{\nabla ( u - v)} \leq \Norms{ \nabla v - y} + \CO \Norms{ \dvg \, y + f } =: M(\nabla v, f, \CO, y) , \qquad \forall y \in \H(\dvg,\O) , \label{eq:majorantestimate}
\end{equation}
where $M$ is called a functional majorant. The global constant $\CO$ represents the smallest possible constant from the Friedrichs' inequality $\Norms{w} \leq\,\CO \Norms{\nabla w}$ which holds for all $w \in \Hoz$. Note that the estimate \eqref{eq:majorantestimate} is sharp: by choosing $y = \nabla u$, the inequality changes into an equality. By this we immediately see that minimizing $M$ with respect to $y$ provides us a way to obtain approximations of the flux $\nabla u$. Since $M$ contains nondifferentiable norm terms, we apply the Young's inequality $(a + b)^2 \leq (1 + \beta) a^2 + (1 + \frac{1}{\beta}) b^2$ valid for all $\beta > 0$ to obtain
\begin{equation*}
\left(\Norms{ \nabla v - y } + \CO \Norms{ {\rm div} y + f }\right)^2
\leq
\left(1+\frac{1}{\beta}\right) \Norms{ \nabla v - y}^2 + (1 + \beta ) \CO^2 \Norms{ \dvg \, y + f  }^2
=:
\mathcal{M}(\nabla v, f, \CO, \beta, y) .
\end{equation*}
The majorant $\mathcal{M}$ arguments $v$ and $f$ are known, and upper bounds of $\CO$ are also known. The parameter $\beta > 0$ and the function $y\in H(\dvg,\O)$ are free parameters. For a fixed value of $\beta$, the majorant represents a quadratic functional in $y$. Global minimization of $\mathcal{M}$ with respect to $y$ results in the following problem for $y$:
\begin{equation} \label{eq:majorantproblem}
	(1+\beta) \CO^2 \!\! \int_\O \dvg \, y \; \dvg \, \phi \, \dx +
	\left(1+\frac{1}{\beta}\right) \!\! \int_\O y \cdot \phi \, \dx =
	- (1+\beta) \CO^2 \!\! \int_\O f \, \dvg \, \phi \, \dx +
	\left(1+\frac{1}{\beta}\right) \!\! \int_\O \nabla v \cdot \phi \, \dx \quad
	\forall \phi \in \H(\dvg,\O)
\end{equation}
On the other hand, for a fixed $y$,
\begin{equation}
\beta= \frac{\Norms{ \nabla v - y } }{ \CO \Norms{ \dvg \, y + f  } }
\label{eq:betaoptimal}
\end{equation}
minimizes $\mathcal{M}$ amongst all $\beta>0$. 
It suggests the following solution algorithm:

\begin{algorithm}[Majorant minimization algorithm] Let $\beta > 0$ be given (for example, set $\beta = 1$).
\item (a) Compute $y$ (using current value of $\beta$) by minimizing the quadratic problem
$\mathcal{M}(\nabla v, f, \CO, \beta, y) \rightarrow \mbox{min} \label{eq:majorantnminimization}$.
\item {
(b) Update $\beta$ (using $y$ calculated in step (a)) from \eqref{eq:betaoptimal}.
If the convergence in $y$ is not achieved then go to step (a).
}
\label{al:majorant_minimization}
\end{algorithm}

We solved the quadratic minimization problem in (a) by discretizing the problem \eqref{eq:majorantproblem} with the linear Raviart-Thomas elements. For this both of the FEM matrices $\mass^\rt$ and $\stiff^\rt$ were needed (see also \cite{Va}).

\begin{example}
In 2D we choose the unit square $\Omega:=(0,1)^2$, and in 3D the unit cube $\Omega:=(0,1)^3$. We choose the bubble function
\begin{equation*}
	u(x) := \prod_{i=1}^d x_i (x_i - 1)
\end{equation*}
as the exact solution in both 2D and 3D. It is clear that $u \in \Hoz$ in both dimensions.
\end{example}

\begin{figure} [h]
\center
\includegraphics[width=0.3\textwidth]{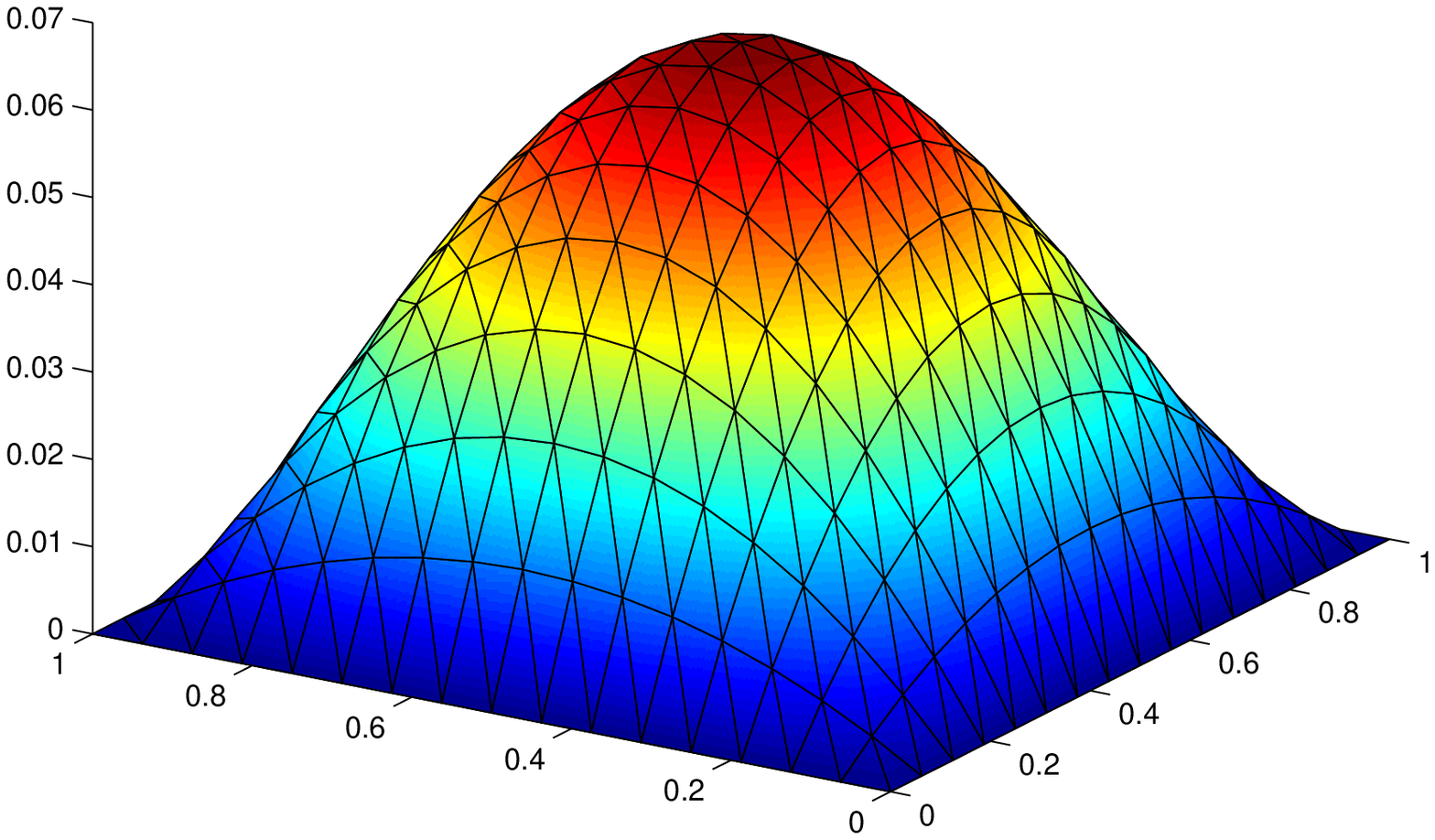}
\includegraphics[width=0.3\textwidth]{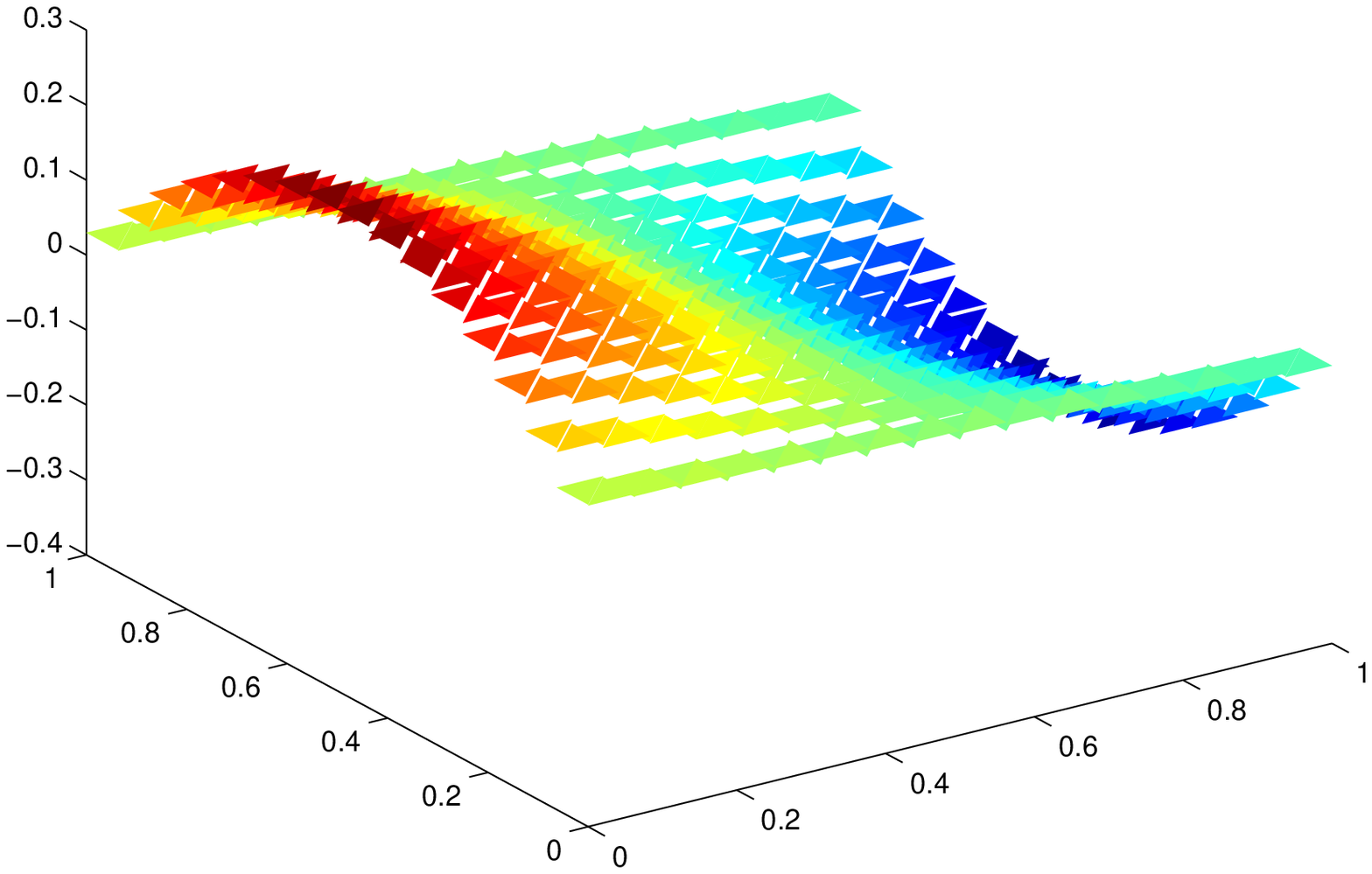}
\includegraphics[width=0.3\textwidth]{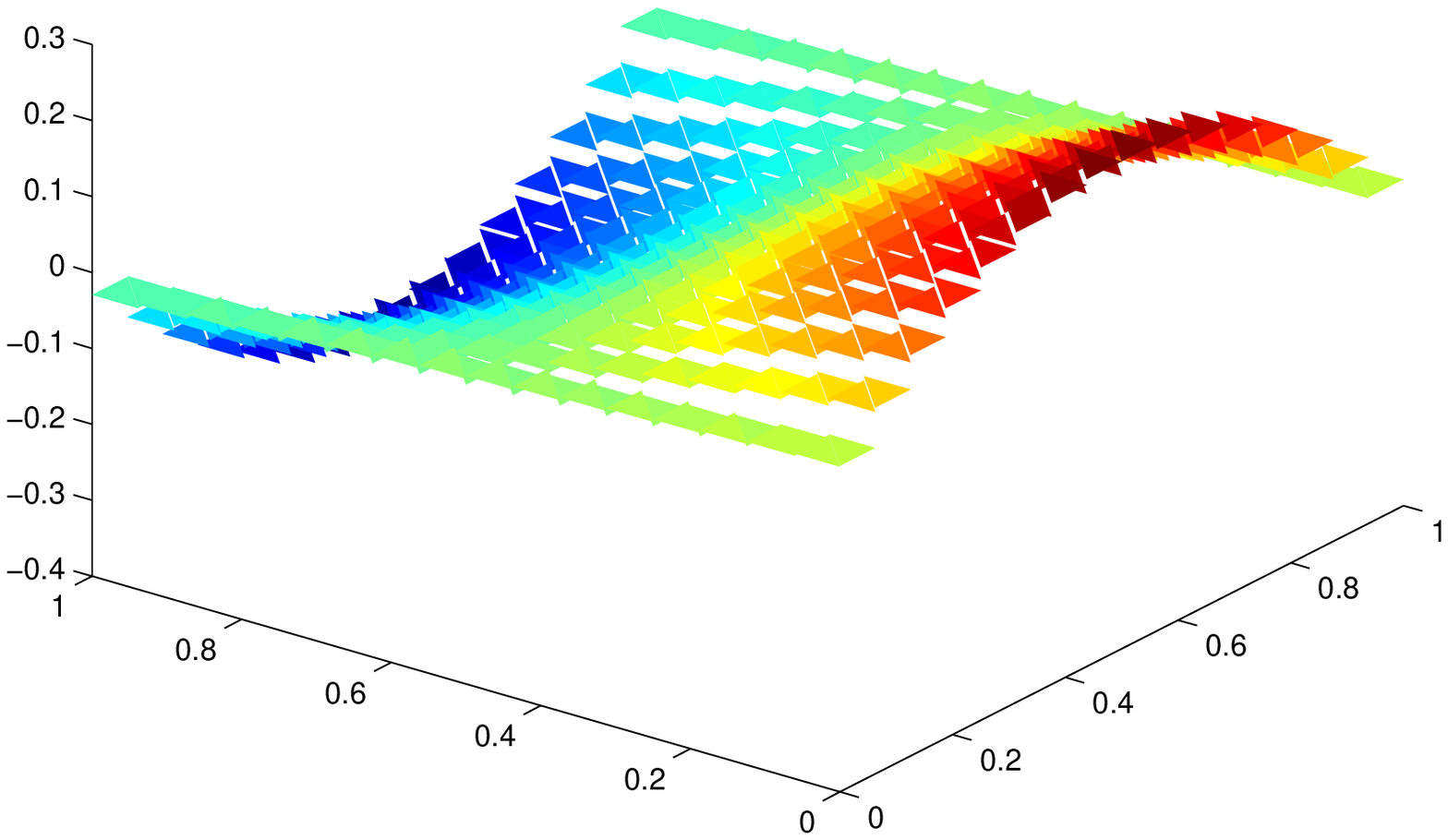}
\caption{Discrete solution $v$ (left), and the flux approximation $y$ first and second components (middle and right) on a mesh with 512 elements.}
\label{fig:poisson}
\end{figure}

On Tables \ref{tab:majorant2d} and \ref{tab:majorant3d} we have calculated the majorant $\mathcal{M}$ values with Algoritm \ref{al:majorant_minimization} for four different meshes in 2D and 3D, respectively. The program code can be found in the folder \verb+/example_majorant/+. The approximation $v$ was calculated with linear nodal finite elements. For measuring the quality of the chosen free parameters $\beta$ and $y$, we have also included the values of the so-called efficiency index $\Ieff := \sqrt{\mathcal{M}}/\Norms{\nabla(u-v)} \ge 1$. The approximation $v$ and flux approximation $y$ of the smallest 2D mesh are depicted in Figure \ref{fig:poisson}. The iterations of Algorihtm \ref{al:majorant_minimization} were stopped if the distance of the previous value of the majorant to the new value (normalized with the previous value) was less than $10^{-4}$.

\begin{table} [t]
\center
{\small
\begin{tabular}{l|l l l|l l l|l l l|l l l}
& \multicolumn{3}{c}{$\#\elems=$ 512} & \multicolumn{3}{|c}{$\#\elems=$ 131 072} & \multicolumn{3}{|c}{$\#\elems=$ \textbf{2} 097 152} & \multicolumn{3}{|c}{$\#\elems=$ \textbf{33} 554 432} \\
Iter & $\beta$ & $\mathcal{M}$ & $\Ieff$ & $\beta$ & $\mathcal{M}$ & $\Ieff$ & $\beta$ & $\mathcal{M}$ & $\Ieff$ & $\beta$ & $\mathcal{M}$ & $\Ieff$ \\
\hline
1 & 1.000 & 0.026203 & 1.72 & 1.000 & 0.001648 & 1.73 & 1.000 & 0.000412 & 1.73 & 1.000 & 0.000103 & 1.73 \\
2 & 3.208 & 0.023159 & 1.52 & 3.294 & 0.001453 & 1.52 & 3.294 & 0.000363 & 1.52 & 3.294 & 0.000091 & 1.52 \\
3 & 3.268 & 0.023159 & 1.52 & 3.294 & 0.001453 & 1.52 & 3.294 & 0.000363 & 1.52 & 3.294 & 0.000091 & 1.52
\end{tabular}
}
\caption{Majorant calculation for four meshes in 2D.}
\label{tab:majorant2d}
\end{table}

\begin{table} [t]
\center
{\small
\begin{tabular}{l|l l l|l l l|l l l|l l l}
& \multicolumn{3}{c}{$\#\elems=$ 10 368} & \multicolumn{3}{|c}{$\#\elems=$ 82 944} & \multicolumn{3}{|c}{$\#\elems=$ 663 552} & \multicolumn{3}{|c}{$\#\elems=$ \textbf{5} 308 416} \\
Iter & $\beta$ & $\mathcal{M}$ & $\Ieff$ & $\beta$ & $\mathcal{M}$ & $\Ieff$ & $\beta$ & $\mathcal{M}$ & $\Ieff$ & $\beta$ & $\mathcal{M}$ & $\Ieff$ \\
\hline
1 & 1.000 & 0.011794 & 1.59 & 1.000 & 0.006176 & 1.59 & 1.000 & 0.003135 & 1.59 & 1.000 & 0.001574 & 1.59 \\
2 & 3.128 & 0.010396 & 1.40 & 3.512 & 0.005379 & 1.39 & 3.655 & 0.002721 & 1.38 & 3.697 & 0.001365 & 1.38 \\
3 & 3.420 & 0.010388 & 1.40 & 3.622 & 0.005379 & 1.39 & 3.687 & 0.002720 & 1.38 & 3.706 & 0.001365 & 1.38 \\
4 & 3.432 & 0.010388 & 1.40
\end{tabular}
}
\caption{Majorant calculation for four meshes in 3D.}
\label{tab:majorant3d}
\end{table}

\subsection{Solving the eddy-current problem using \nedelec elements}
We split the boundary into two parts: $\partial\Omega := \Gamma_D \cup \Gamma_N$ such that $\Gamma_D \cap \Gamma_N = \emptyset$. The 2D eddy-current problem reads as
\begin{align*}
	\curlv \, \mu^{-1} \curl \, E + \kappa E & = F && \mbox{in } \O , \\
	E \times n & = 0 && \mbox{in } \Gamma_D , \\
	\mu^{-1} \curl \, E & = 0 && \mbox{in } \Gamma_N , 
\end{align*}
for $E \in \Hoc := \{ v \in \H(\curl,\O) \mid v \times n = 0 \; \textrm{on} \; \Gamma_D \}$, where $n$ denotes the outward unit normal to the boundary $\partial\Omega$. Here the right hand side $F \in \Ltwo(\O,\R^2)$, and the positive material parameters $\mu,\kappa \in L^{\infty}(\Omega)$ are given. The exact solution $E$ is sought from the weak formulation
\begin{equation} \label{eq:eddygeneral}
	\int_\O \mu^{-1} \curl \, E \; \curl \, w \, \dx + \int_\O \kappa E \cdot w \, \dx = \int_\O F \cdot w \, \dx \qquad \forall w \in \Hoc .
\end{equation}

\begin{example}[\cite{equality}]
We choose the unit square $\Omega:=(0,1)^2$ with $\kappa=\mu=1$. We split the domain in two parts across the diagonal, $\Omega_1:=\{ x\in\Omega \mid x_1 > x_2\}$ and $\Omega_2:=\Omega\backslash\up{\Omega}_1$, in order to define the following discontinuous exact solution:
\begin{equation*}
	\restr{E}{\Omega_1}(x):=
	\begin{pmatrix}
	\sin(2\pi x_1)+2\pi\cos(2\pi x_1)(x_1-x_2)\\
	\sin\big((x_1-x_2)^2(x_1-1)^2x_2\big) - \sin(2\pi x_1)
	\end{pmatrix},\quad
	\restr{E}{\Omega_2}(x):=0 .
\end{equation*}
Since on $\Gamma_/ := \{x \in \Omega \mid x_1=x_2\}$ we have
\begin{equation*}
	\restr{E}{\Omega_1} \times n =
	\frac{1}{\sqrt{2}} \left( 2\pi\cos(2\pi x_1)(x_1-x_2) +
	\sin\big((x_1-x_2)^2(x_1-1)^2x_2\big)
	\right) , \quad
	\restr{E}{\Omega_2} \times n = 0 ,
\end{equation*}
we see that $\restr{E}{\Omega_1} \times n = 0$ on $\Gamma_/$. We conclude that the tangential component is continuous on $\Gamma_/$, so $E$ belongs to $\H(\curl,\O)$. Moreover,
\begin{equation*}
	\restr{\curl \, E}{\Omega_1} =
	2 x_2 (x_1-x_2) (x_1-1) (2x_1-x_2-1) \cos\big(x_2(x_1 - x_2)^2 (x_1 - 1)^2\big) , \quad
	\restr{\curl \, E}{\Omega_2} = 0 ,
\end{equation*}
and clearly $\restr{\curl \, E}{\Omega_1} = 0$ on $\Gamma_/$, i.e., $\curl \, E$ is continuous on $\Gamma_/$. Also, it is easy to see that $\curl \, E$ vanishes on the whole boundary, so it belongs to $\Hoz$. Thus, the exact solution satisfies zero Neumann boundary condition on the whole boundary, i.e., $\Gamma_D=\emptyset$ and $\Gamma_N=\partial\Omega$.
\end{example}

\begin{figure}
\center
\includegraphics[width=0.3\textwidth]{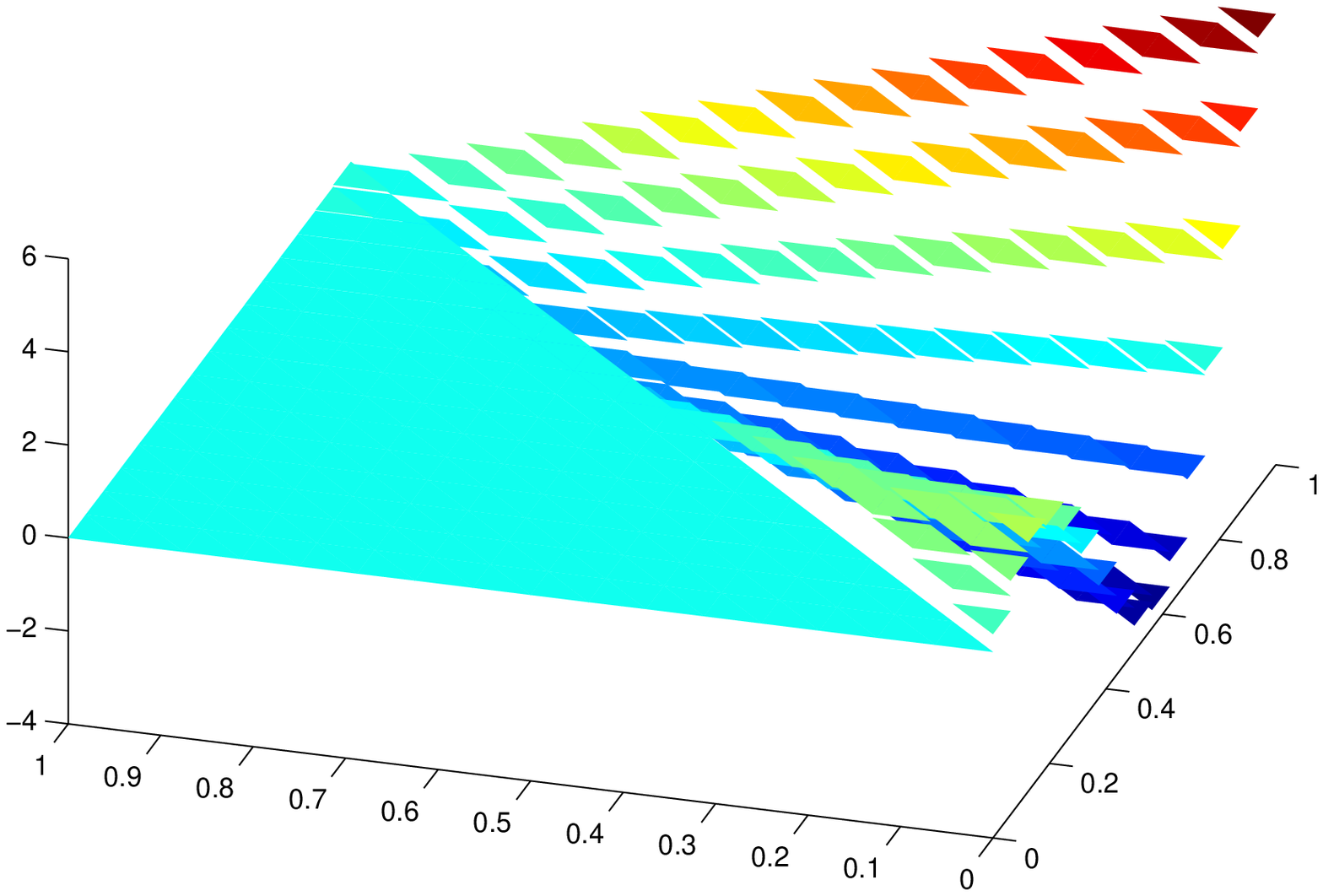}
\includegraphics[width=0.3\textwidth]{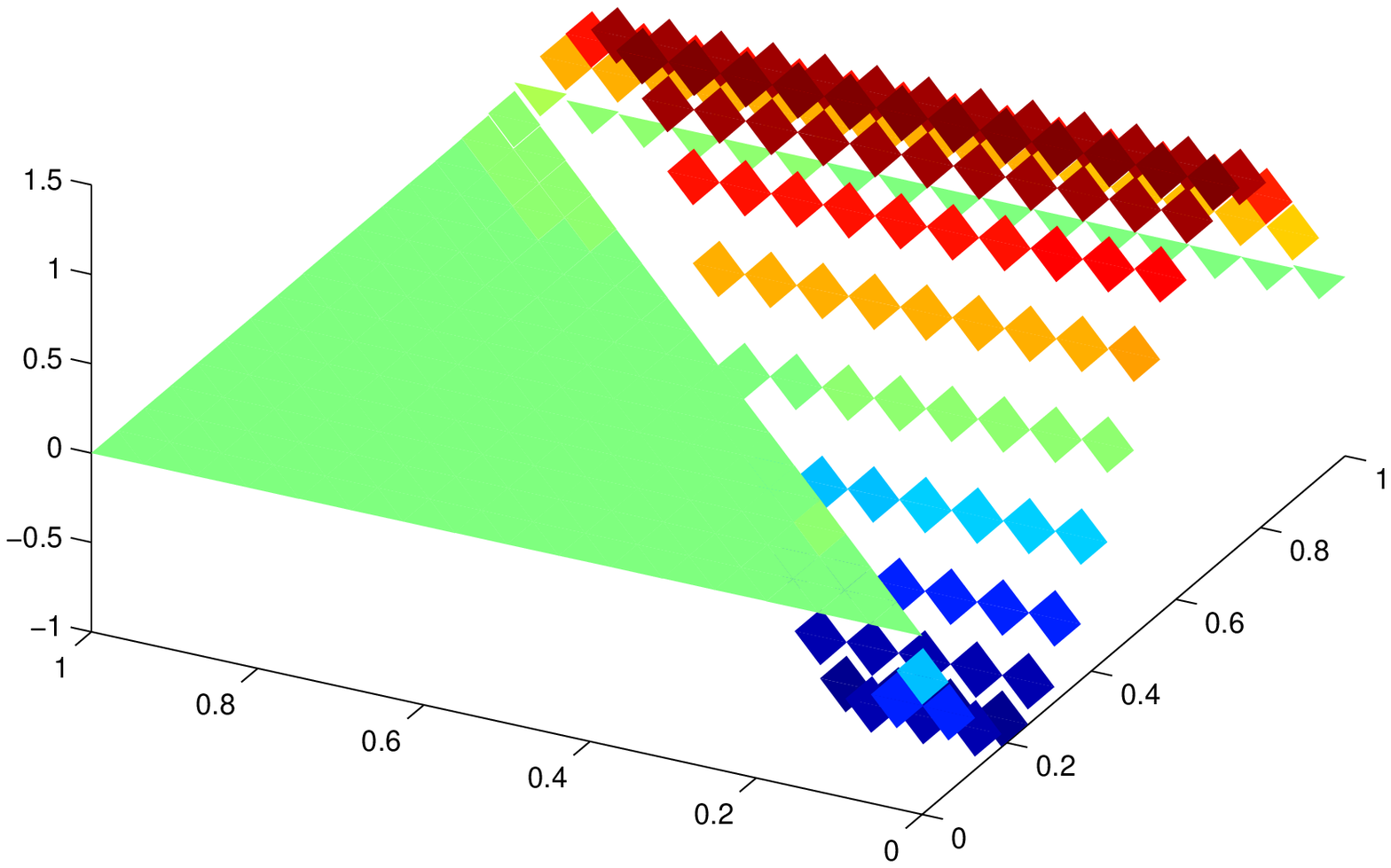}
\includegraphics[width=0.3\textwidth]{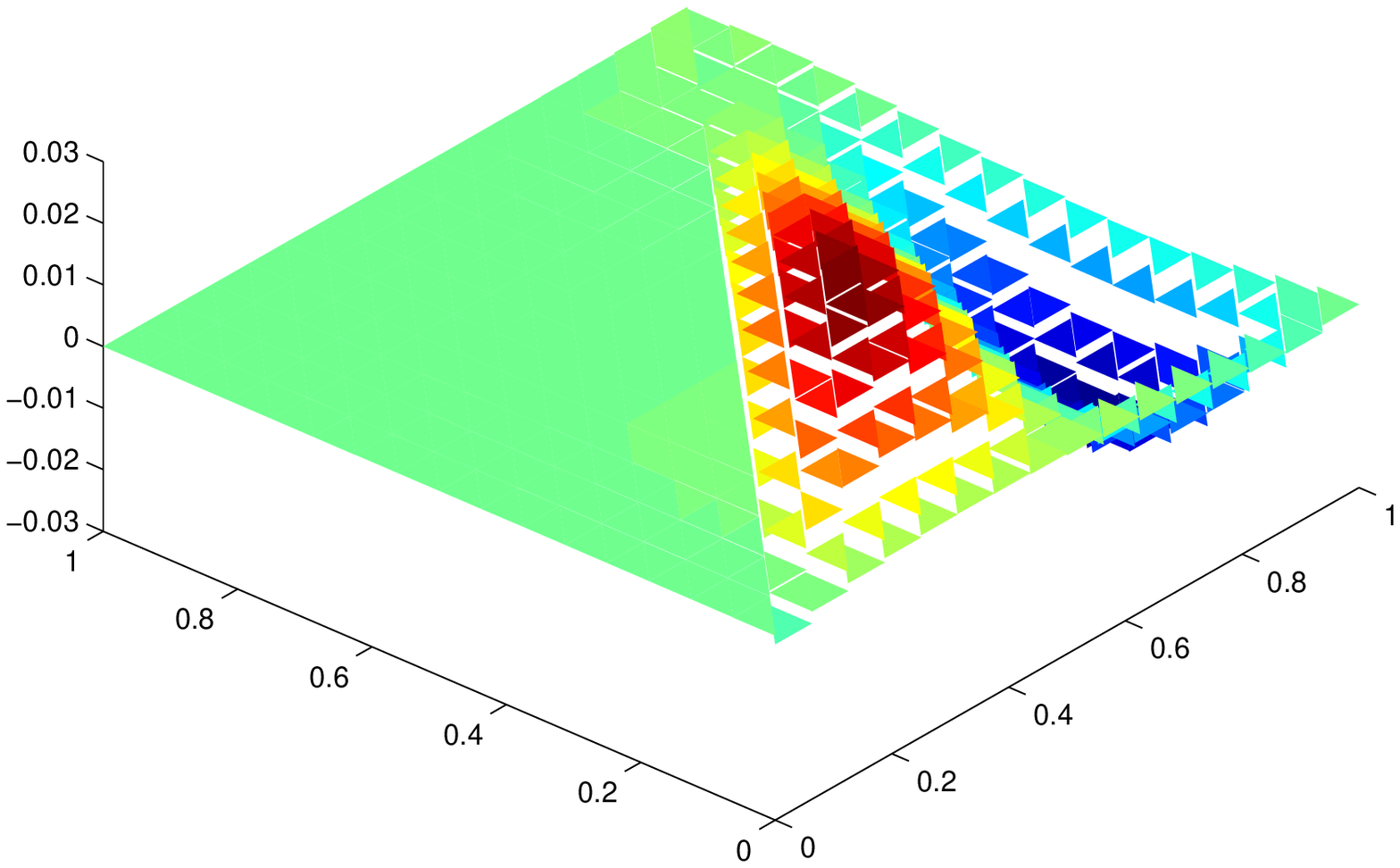}
\caption{Discrete solution $v$ first and second components (left and middle), and $\curl \, v$ (right) on a mesh with 512 elements.}
\label{fig:eddy}
\end{figure}

We denote by $v$ an approximation of the exact solution $E$ of \eqref{eq:eddygeneral}. In the discretization of \eqref{eq:eddygeneral} we need both the mass and stiffness matrices $\mass^\ned$ and $\stiff^\ned$. We see from Figure \ref{fig:eddy} that the 2D \nedelec element catches the normal discontinuity on the diagonal line $\Gamma_/$. In Table \ref{tab:eddy} we show how the error measured in the $\H(\curl,\O)$ -norm decreases as the mesh is uniformly refined. The program code can be found in the folder \verb+/example_eddycurrent/+.

\begin{table}
\center
{\small
\begin{tabular}{r r | c}
$\#\elems$ & $\#\edges$ & $\sqrt{\Norms{E - v}^2 + \Norms{\curl(E - v)}^2}$ \\
\hline
32 768 & 49 408 & 2.358185e-02 \\
131 072 & 197 120 & 1.179151e-02 \\
524 288 & 787 456 & 5.895834e-03 \\
\textbf{2} 097 152 & \textbf{3} 147 776 & 2.947927e-03 \\
\textbf{8} 388 608 & \textbf{12} 587 008 & 1.473965e-03 \\
\textbf{33} 554 432 & \textbf{50} 339 840 & 7.369826e-04 \\
\end{tabular}
}
\caption{Exact energy errors of approximations of the 2D eddy-current problem on uniformly refined meshes.}
\label{tab:eddy}
\end{table}

\begin{acknow}{5}
The work of the first author was supported by the V\"ais\"al\"a Foundation of the Finnish Academy of Science and Letters. The second author acknowledges the support of the project GA13-18652S (GA CR).
\end{acknow}

\bibliographystyle{plain}
\bibliography{Anjam_Valdman_biblio}

\begin{thebibliography}{10}

\bibitem{equality}
I.~Anjam and D.~Pauly.
\newblock Functional a posteriori error equalities for conforming mixed
  approximations of elliptic problems.
\newblock Preprint, 2014.
\newblock URL: \url{http://arxiv.org/abs/1403.2560} (accessed 9.1.2015).

\bibitem{BaCa}
C.~Bahriawati and C.~Carstensen.
\newblock Three {MATLAB} implementations of the lowest-order {Raviart-Thomas}
  {MFEM} with a posteriori error control.
\newblock {\em CMAM}, 5(4):333--361, 2005.

\bibitem{BrFo}
F.~Brezzi and M.~Fortin.
\newblock {\em Mixed and hybrid finite element methods}, volume~15 of {\em
  Springer Series in Computational Mathematics}.
\newblock Springer-Verlag, New York, 1991.

\bibitem{Chen}
L.~Chen.
\newblock {iFEM}: an integrated finite element method package in {MATLAB}.
\newblock Technical report, University of California at Irvine, 2009.
\newblock URL: \url{http://math.uci.edu/~chenlong/programming.html} (accessed
  9.1.2015).

\bibitem{CuJaSc}
F.~Cuvelier, C.~Japhet, and G.~Scarella.
\newblock An efficient way to perform the assembly of finite element matrices
  in {Matlab} and {Octave}.
\newblock Preprint, 2013.
\newblock URL: \url{http://arxiv.org/abs/1305.3122} (accessed 9.1.2015).

\bibitem{Du}
D.~A. Dunavant.
\newblock High degree efficient symmetrical gaussian quadrature rules for the
  triangle.
\newblock {\em Int. J. Num. Methods Eng.}, 21:1129--1148, 1985.

\bibitem{FuPrWi}
S.~Funken, D.~Praetorius, and P.~Wissgott.
\newblock Efficient implementation of adaptive {P1-FEM} in {MATLAB}.
\newblock {\em Comput. Methods Appl. Math.}, 11:460--490, 2011.

\bibitem{HaJu}
A.~Hannukainen and M.~Juntunen.
\newblock Implementing the finite element assembly in interpreted languages.
\newblock Preprint, Aalto University, 2012.

\bibitem{Ned}
J.~C. N\'ed\'elec.
\newblock Mixed finite elements in {$\mathbb{R}^3$}.
\newblock {\em Numerische Matematik}, 35:315--341, 1980.

\bibitem{Re2004}
P.~Neittaanm\"aki and S.~Repin.
\newblock {\em Reliable methods for computer simulation. Error control and a
  posteriori estimates}, volume~33 of {\em Studies in Mathematics and its
  Applications}.
\newblock Elsevier, Amsterdam, 2004.

\bibitem{RaVa}
T.~Rahman and J.~Valdman.
\newblock Fast {MATLAB} assembly of {FEM} matrices in {2D} and {3D}: Nodal
  elements.
\newblock {\em Applied Mathematics and Computation}, 219:7151--7158, 2013.

\bibitem{RT}
P.~A. Raviart and J.~M. Thomas.
\newblock A mixed finite element for second order elliptic problems.
\newblock In I.~Galligani and E.~Magenes, editors, {\em Mathematical Aspects of
  Finite Element Methods}, pages 292--315. Springer-Verlag, New York, 1977.

\bibitem{Schneebeli}
A.~Schneebeli.
\newblock An {$H(\curl;\Omega)$}-conforming {FEM}: {N\'ed\'elec's} elements of
  first type.
\newblock Technical report, 2003.
\newblock URL: \url{http://www.dealii.org/reports/nedelec/nedelec.pdf}
  (accessed 9.1.2015).

\bibitem{Sch}
J.~Sh\"oberl.
\newblock {C}++11 implementation of finite elements in {NGS}olve.
\newblock ASC Report 30/2014, Institute for Analysis and Scientific Computing,
  Vienna University of Technology, 2014.

\bibitem{So2}
P.~{\v S}ol{\'\i}n, L.~Korous, and P.~Kus.
\newblock Hermes2{D}, a {C}++ library for rapid development of adaptive
  hp-{FEM} and hp-{DG} solvers.
\newblock {\em Journal of Computational and Applied Mathematics}, 270:152--165,
  2014.

\bibitem{So}
P.~{\v S}ol{\'\i}n, K.~Segeth, and I.~Dole{\v z}el.
\newblock {\em Higher-order finite element methods}, volume~41 of {\em Studies
  in Advanced Mathematics}.
\newblock Chapman and Hall/CRC, Boca Raton, Florida, 2003.

\bibitem{Va}
J.~Valdman.
\newblock Minimization of functional majorant in a posteriori error analysis
  based on {H(div)} multigrid-preconditioned cg method.
\newblock {\em Advances in Numerical Analysis}, vol. 2009, 2009.
\newblock Article ID 164519.

\bibitem{ZhCuLi}
L.~Zhang, T.~Cui, and H.~Liu.
\newblock A set of symmetric quadrature rules on triangles and tetrahedra.
\newblock {\em J. Comp. Math.}, 26(3):1--16, 2008.

\end{thebibliography}

\end{document}